\documentclass[11pt,a4]{article}

\usepackage{graphicx}
\usepackage[font=footnotesize,labelfont=bf]{caption}
\usepackage{subfigure}
\usepackage{hyperref}
\usepackage{color,amsmath,enumerate}
\usepackage[top=2.5cm, bottom=2.5cm, left=1in, right=1in]{geometry}
\numberwithin{equation}{section}

\begin{document}
\bibliographystyle{apalike}

\title{The Hierarchical Spectral Merger algorithm: A New Time Series Clustering 
Procedure}

\author{Carolina Eu\'an \footnote{Centro de Investigaci\'on en 
Matem\'aticas, Guanajuato, Gto, M\'exico.} \footnotemark,
Hernando Ombao \footnote{Department of Statistics, University of 
California, Irvine, USA.} 
\addtocounter{footnote}{-2}\addtocounter{Hfootnote}{-2}\footnote{UC Irvine 
Space-Time Modeling Group.}, Joaqu\'in Ortega 
\addtocounter{footnote}{-2}\addtocounter{Hfootnote}{-2}\footnotemark }

\date{}
\maketitle

\begin{abstract}
We present a new method for time series clustering which we call the 
Hierarchical Spectral Merger (HSM) method. This 
procedure is based on the spectral theory of time series and identifies series 
that
share similar oscillations or waveforms. 
The extent of similarity between a pair of time series is measured using 
the total variation distance between their estimated spectral densities.
At each step of the algorithm, every time two clusters merge, a new spectral density is estimated using the whole information present in both clusters, which is representative of all the series in the new cluster. 
The method is implemented in an R package \textit{HSMClust}. We  present two 
applications of the HSM method, one to data coming from 
wave-height measurements in oceanography and the other to electroencefalogram 
(EEG) data. 
\end{abstract}

\vspace{.5in}


\section{Introduction}

The subject of time series clustering is an active research area with 
applications in many fields. Finding similarity between time series frequently
plays a central role in many applications. In fact, time series clustering 
problems arise in a natural way in a
wide variety of fields, including economics, finance, medicine, ecology, 
environmental studies, engineering, and many others.  A recent work, developed 
by \cite{Krafty16}, uses a clustering model to develop a discriminant 
analysis of stochastic cepstra. Time series clustering 
is, in general, not an easy task due to the potential complexity of the data 
and the difficulty of defining an 
adequate notion of similarity between time series.

In \cite{Liao05}, and subsequently in \cite{Caiado15}, there are three 
approaches to time
series clustering: methods based on the comparison of raw data,
methods based on parameters from models adjusted to the data, and 
feature-based methods where the similarity between time series is
measured through features extracted from the data. 
The first clustering approach compares of raw data and may not be 
computationally scalable for long time series. The second 
approach, based on parameters, is one of the most frequently used. However, it 
has the limitation of requiring a parametric model and this might suffer from 
model misspecification. 
\cite{Vilar14} present an R package (TSclust) for time series
clustering with a wide variety of alternative procedures. 

Here, we consider the problem of clustering stationary time series and our 
proposal is based on using the spectral density as the central feature  for 
classification purposes. To build a clustering method one needs to measure the 
similarity between the spectral densities. Our method uses the total 
variation (TV) 
distance as a measure of dissimilarity, which was proposed in 
\cite{Alvarez15}. 
This distance requires the use of normalized spectral densities, which is 
equivalent to standardizing the time series so that their variances are equal 
to 
one. Thus, it is the distribution of the power across different frequencies of 
the time series that is considered the fundamental feature for clustering 
purposes rather than the magnitude of the oscillations.

In \cite{Alvarez15}, the TV distance was used to build a dissimilarity matrix 
consisting of the distances between all the spectral densities, which was then 
fed into a classical hierarchical agglomerative algorithm with the complete and 
average linkage functions. The method introduced in this work which we call 
the 
Hierarchical Spectral Merger (HSM) method, is a new clustering 
algorithm that takes advantage of the spectral theory of time series.
The key difference with classical hierarchical agglomerative clustering  is 
that every time a cluster is updated, a new representative (new estimated 
spectral density) is computed. 
Each time two clusters are joined in the HSM procedure, the 
information available in all the series belonging to both clusters is merged to 
produce a new estimate of the common spectral density for the new cluster. 
The proposed approach is appealing because it is intuitive and the updated 
spectral estimates are 
smoother, less noisy and hence give better estimates of the TV distance. 
Thus, every time two clusters merge, the dissimilarity matrix reduces its size 
in one unit. In contrast, for the classical hierarchical agglomerative algorithm 
the dissimilarity matrix is the same throughout the procedure, and the distances 
between clusters at each step are calculated using linear combinations of the 
individual distances; the linear combination used depends on the linkage 
function that is chosen (single, complete, average, Ward, etc.). These methods 
are based on geometrical ideas which, in some cases, may not be meaningful for 
clustering time series, since a linear combination of the individual distances 
may not have a clear meaning in terms of the spectral densities.

We will present two applications of the HSM method: one to data coming from 
wave-height measurements in oceanography and the other to electroencephalogram 
(EEG) data. Some of the numerical experiments described in Section 3 are 
related 
to these applications. 

The rest of the paper is organized as follows: Section \ref{sec2} describes the 
Hierarchical Spectral Merger procedure. Section 3 presents some numerical 
experiments which compare the HSM method with some of the existing algorithms 
and 
considers the problem of deciding how many clusters there are in a given 
sample. 
Finally, Section 4 gives some examples of the use of the HSM algorithm. The 
paper ends with some discussion of the results and some ideas about future work.
\section{HSM: A Method for Time Series Clustering } \label{sec2}

Our goal is to develop a method that finds groups or clusters that represent 
spectrally synchronized time series. The algorithm we introduce, known as the 
Hierarchical Spectral Merger (HSM) method, uses the TV distance as a 
dissimilarity measure and proposes a new clustering procedure. This
algorithm is a modification of the usual agglomerative hierarchical 
procedure, taking advantage of the spectral point of view for the analysis of 
time series.

The first question when building a clustering algorithm is how to measure the 
dissimilarity between the objects one is considering. In our case, this amounts 
to measuring the dissimilarity between the spectral densities estimated from 
the original time series, for which we use the TV distance between the 
normalized spectra. 
In what follows, section \ref{sec2-1} presents the TV distance while 
section \ref{sec2-2} describes in detail the HSM method.

\subsection{Total Variation (TV) Distance}\label{sec2-1}
In general, the TV distance can be defined for any pair of probability measures 
that 
live on the same $\sigma$-algebra of sets. We will be focus here in the case 
where these probability measures are defined on the real line and have density 
functions with respect to the 
Lebesgue measure.
The TV distance between two probability densities, $f$ and $g$, is defined as
\begin{eqnarray}\label{TVd_1}
d_{TV}(f,g)=1-\int\min\{f(\omega),g(\omega)\}\, \mbox{d}\omega.
\end{eqnarray}

\begin{figure}
\centering
\includegraphics[scale=.35]{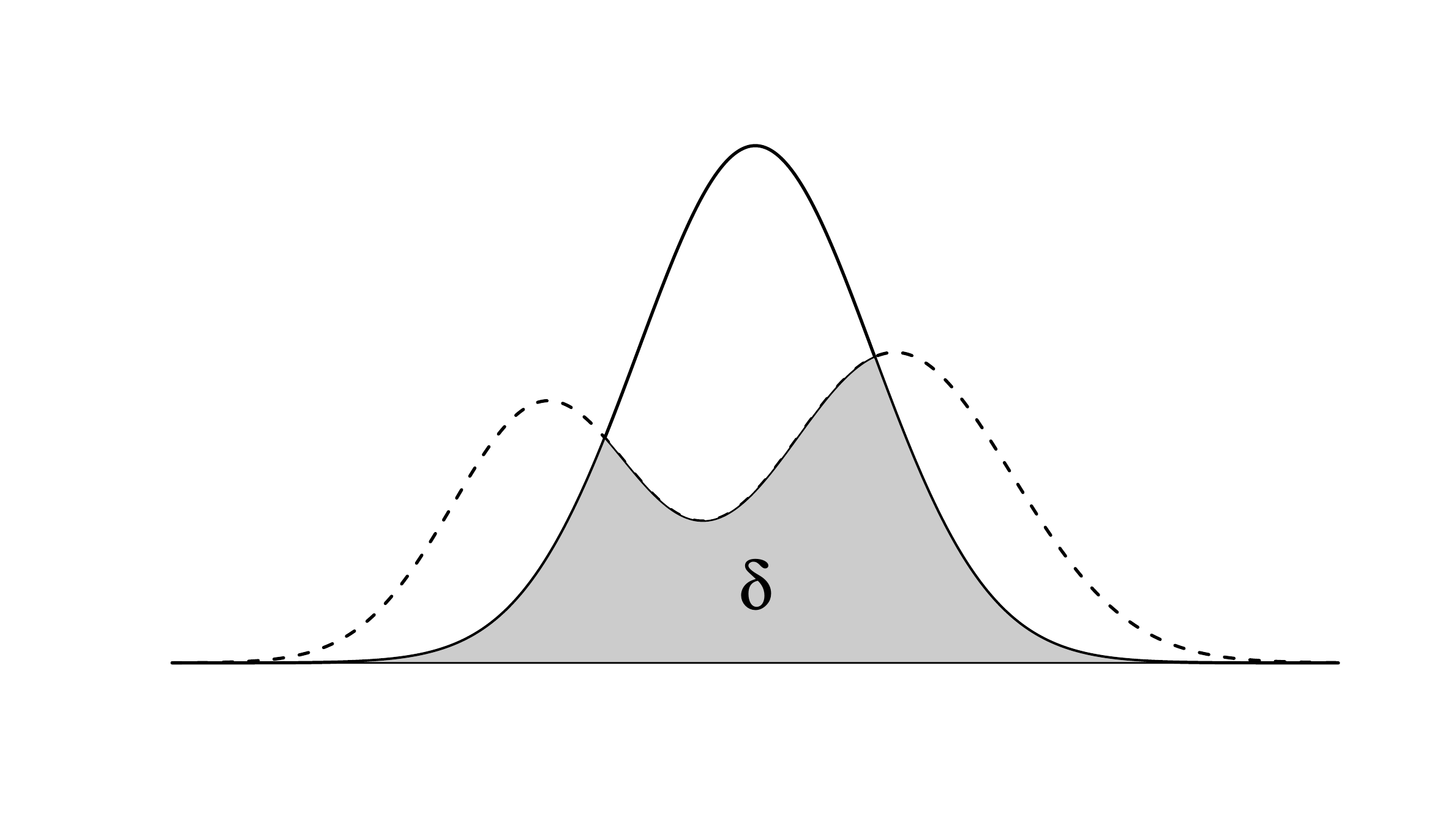}
\caption{The total variation distance measures the similarity between two 
densities. The shaded area represents the common area below the two densities, 
which we denote as $\delta$.}\label{Fig1}
\end{figure}

Equation (\ref{TVd_1}) suggests a graphical interpretation of the TV distance. 
If $f$ and $g$ are probability density functions and $d_{TV}(f,g) = 1-\delta$ 
then $\int\min\{f(\omega),g(\omega)\}\, \mbox{d}\omega = \delta$ and this means 
that the common area below the two densities is equal to $\delta$, which 
corresponds to the shaded area in figure \ref{Fig1}. When the area shared by 
the 
two densities increases then the TV distance decreases. 

Since spectral densities are not probability densities, they have to be 
normalized by dividing the estimated spectral density by the sample variance 
$\widehat\gamma(0)$, since $\widehat{\gamma}(0) = \int \widetilde{f}(\omega) 
d\omega$. Thus, we denote 
$\widehat{f}(\omega)=\widetilde{f}(\omega)/\widehat\gamma(0)$ to be the 
normalized 
estimated spectral density.

In comparison with other similarity measures, the TV distance has some 
desirable 
properties. (1) The TV distance is a pseudo metric. It satisfies the symmetry 
condition 
and the 
triangle 
inequality, which are two reasonable properties expected from a similarity 
measure. In this sense, the TV distance may be a better choice than the 
Kullback-Leibler (KL)
divergence (although one can symmetrize the KL divergence). (2) The TV distance 
is bounded, $0 \leq d_{TV}(f,g)\leq 1$ and can 
be 
interpreted in terms of the common area between two densities. Having a bounded 
range ($[0,1]$) is a desirable feature, since this gives a very intuitive sense 
to the values attained by the similarity measure. A value near $0$ means that 
the densities are similar while a value near $1$ indicates they are highly 
dissimilar. In contrast, both the $L^2$ distance and the Kullback-Leibler 
divergence are not bounded above and thus lack this intuitive interpretation.

\subsection{The Hierarchical Spectral Merger (HSM) Algorithm}\label{sec2-2}
There are two general families of clustering algorithms: partitioning and 
hierarchical. Among partitioning algorithms, the K-means and K-medoids are the 
most commonly used. \cite{Maharaj12} proposed a K-means fuzzy clustering 
method based on wavelets coefficients and made a comparison with other K-means 
algorithms. For the hierarchical clustering algorithms, the typical 
examples are agglomerative with single-linkage or complete-linkage 
[\cite{Xu05}]. Storage and computational properties of the hierarchical 
clustering methods are reviewed in \cite{Fionn15}.

The hierarchical spectral merger algorithm has two versions:
the first, known as \textit{single version}, updates the spectral 
estimate of the cluster from a concatenation of the time series, and the 
second, known as \textit{average version}, updates the spectral estimate of the 
cluster from a weighted average of the spectral estimate obtained from each 
signal in the cluster.
\bigskip

\textbf{\textit{Hierarchical Spectral Merger Algorithm.}}
Let $\{X_i=(X_i(1),\ldots,X_i(T)), i=1,\ldots,N\}$ be a set of time series. The 
procedure starts with $N$ clusters, each cluster being a single time series.\\
 \noindent
 \textbf{Step 1.} Suppose there are k clusters. For each cluster, 
estimate the spectral density (using some smoothing or averaging method) and 
represent each cluster by a common normalized spectral 
density $\widehat{f}_j(\omega)$, $j=1,\ldots,k$.\\
 \textbf{Step 2.} Compute the TV distance between these $k$ spectral 
densities.\\
 \textbf{Step 3.} Find the two clusters that have smallest TV distance. \\
 \textbf{Step 4.} Merge the time series in the two clusters with the smallest 
TV 
distance and replace the two clusters with the newly combined single cluster.\\
\textbf{Step 5.} Repeat Steps 1-4 until there is only one cluster left.

\bigskip
In Step 1, at the first iteration the spectral density for each time series is 
initially estimated using the smoothed periodogram. In our case we used a lag 
window estimator with a Parzen window. In further iterations, however, the way 
the common spectral density is obtained depends on the version of the 
algorithm.
When two clusters merge, there are two options, either (1) for the single 
version, we concatenate the signals in both clusters and compute the smoothed 
periodogram with the concatenated signal; or (2) for the average version, we 
take the weighted average over all the estimated spectra 
for each signal in the new cluster as the new estimated spectra. 

\begin{table}
\begin{center}
\footnotesize
 \begin{tabular}{l}
 \hline
 \textbf{Algorithm:}\\
 \hline \hline \\
\begin{minipage}{\linewidth}
 \begin{enumerate}\footnotesize
 \item  Initial clusters: $\mathbf{C}=\{C_i\}$,  $C_i=X_i$, $i=1,\ldots,N$.\\
 Dissimilarity matrix entry between clusters $i$ and $j$, \\
 $$D_{ij}=d(C_i,C_j):=d_{TV}(\widehat{f}_i,\widehat{f}_j),$$ 
 TV distance between the corresponding estimated normalized spectral densities 
$\widehat{f}_i$ using the signals in each cluster.
 
 \item \textbf{for} $k$ $in$ $1: N-1$
 \item \hspace{.3cm}$\displaystyle (i_{k},j_{k})=\arg\!\min_{ij} D_{ij};~ 
\displaystyle min_k=\min_{ij} D_{ij}$ \hspace{1.2cm} $\#$Find the closest 
clusters
 \item \hspace{.3cm}$C_{new}=C_{i_{k}} \cup C_{j_{k}}$\hspace{4.55cm}

        $\#$Join the closest clusters
 \item \hspace{.3cm}$D^{new}=D\setminus \{D_{i_{k} .} \cup D_{j_{k} .} \cup 
D_{. 
i_{k}} \cup D_{. j_{k}}\}$ \hspace{1.05cm}  $\#$Delete rows and columns 
$i_{k},j_{k}$
 \item \hspace{.3cm}\textbf{for} $j$ $in$ $1: N-k-1$
 \item \hspace{.6cm}$D_{(N-k)j}^{new} {= D_{j(N-k)}^{new}=}d_{TV}(C_{new},C_j)$ 
\hspace{1.3cm} 
$\#$Compute new distances
 \item \hspace{.3cm}\textbf{end}
 \item \hspace{.3cm}$D = D^{new}; ~ \displaystyle \mathbf{C}= 
\left(\mathbf{C}\setminus \{C_{i_k},C_{j_k}\} \right) \cup C_{new}$ 
\hspace{.9cm} 
   $\#$New matrix $D$ and new clusters
 \item \textbf{end}
\end{enumerate}
\end{minipage}
\\
\\
\hline
 \end{tabular}
\caption{Hierarchical Spectral Merger Algorithm proposed using the Total 
Variation 
distance and the estimated spectra.}\label{Algo}
\end{center}
 \end{table}

\textit{\bf{Remarks.}} (1) The value in Step 3 represents the ``cost'' of 
joining 
two clusters, i.e., having $k-1$ clusters versus $k$ clusters.
If a significantly large value is observed, then it may be reasonable to choose 
$k$ clusters instead of $k-1$.
(2) Both versions of the algorithm compute the TV distance between the new and 
the old 
clusters based on updated estimated spectral densities, which is the main 
difference with 
classical hierarchical algorithms. While a hierarchical algorithm has a 
dissimilarity matrix of size $N\times N$ during the whole procedure, the 
proposed method reduces this size to $(N-k) \times (N-k)$ at the $k$-th 
iteration. Table \ref{Algo} gives a summary of the hierarchical merger 
algorithm.

\begin{figure}
\centering
\subfigure[ 
\label{F36a}]{\includegraphics[scale=.5]{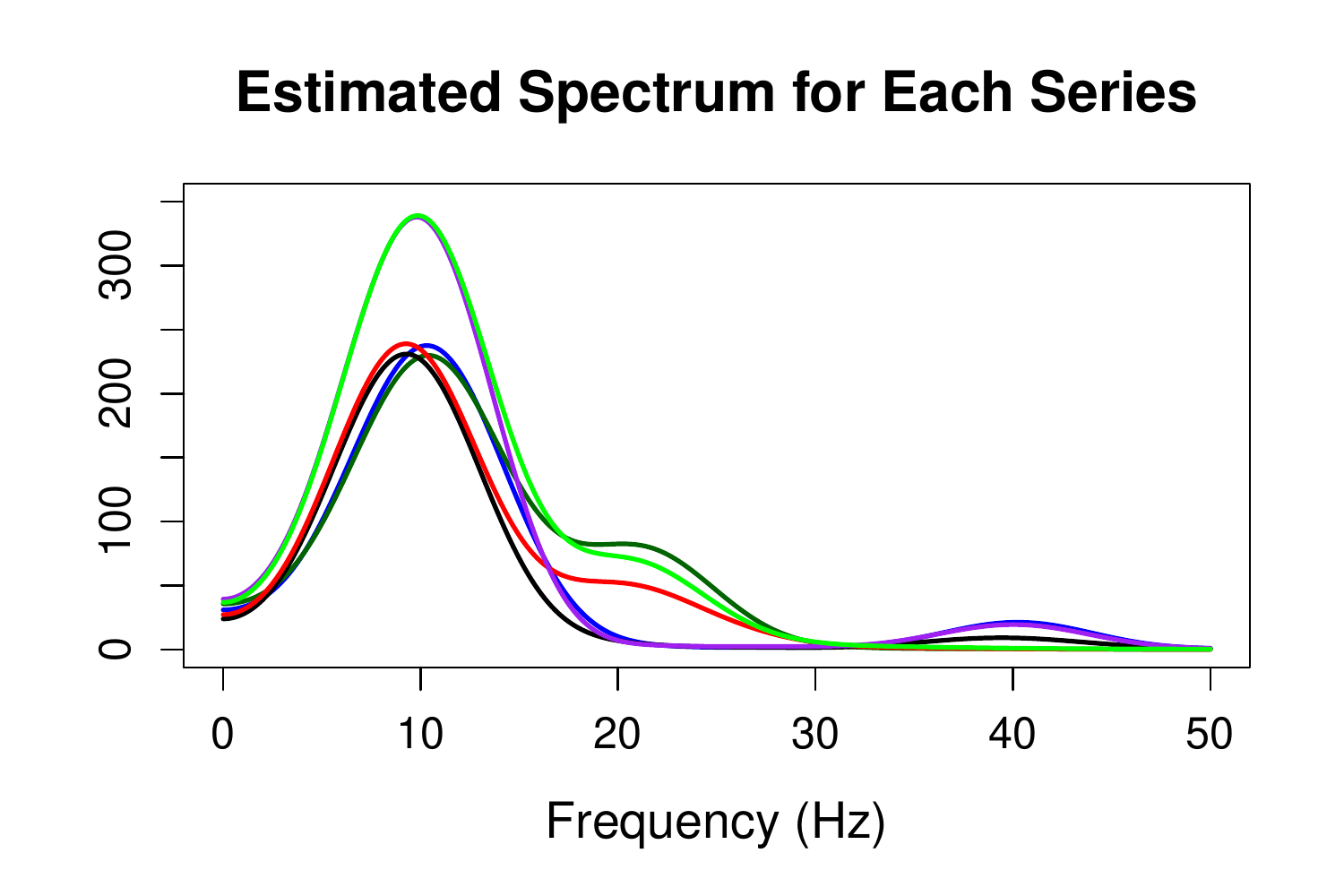}}
\subfigure[ 
\label{F36b}]{\includegraphics[scale=.5]{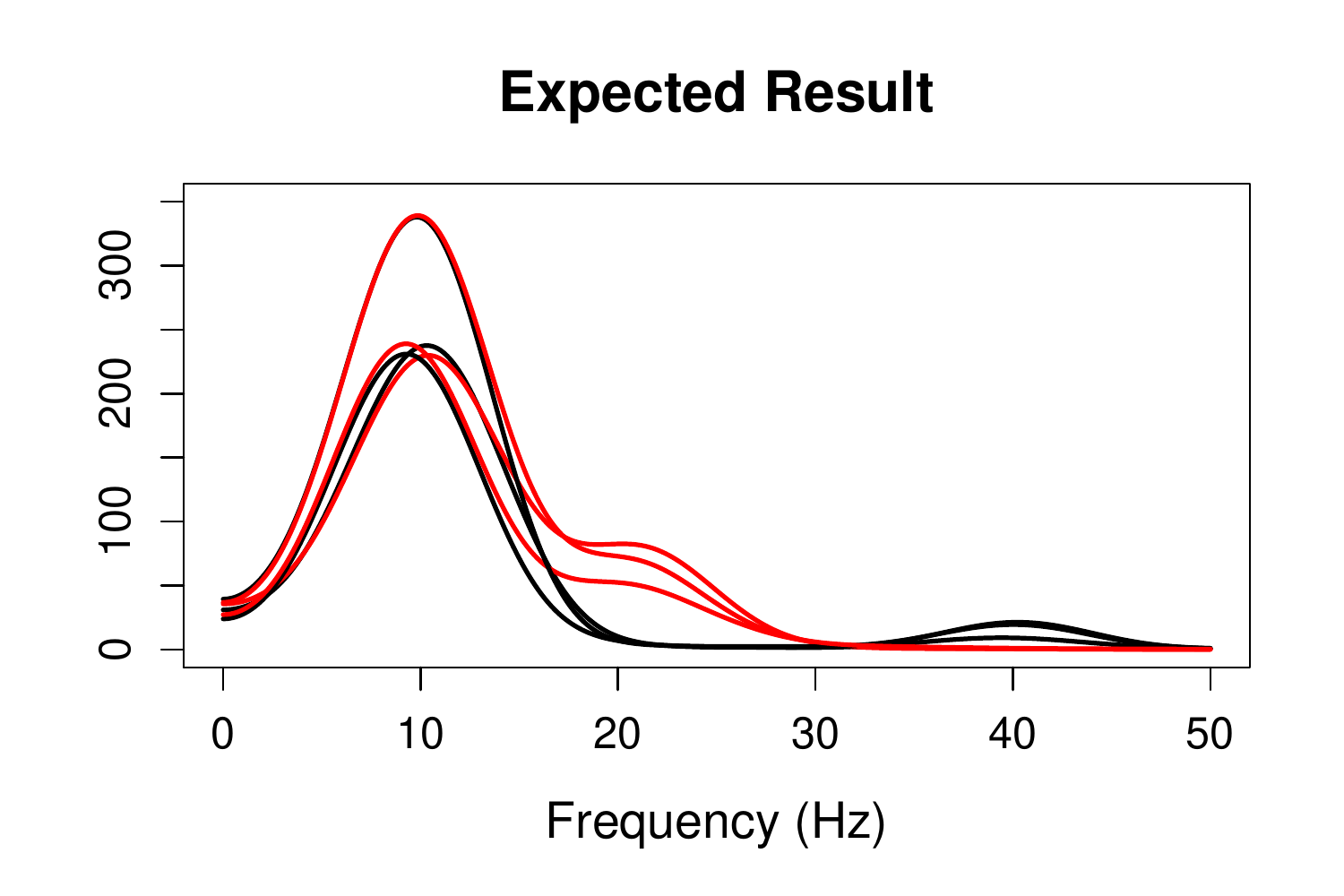}}
\caption{Estimated spectra. (a.) Different colors correspond to different time 
series, (b.) Red spectra are from realizations of the AR(2) model with activity 
at alpha (8-12 
Hz) and beta (12-30 Hz) bands and black spectra are from realizations of the 
AR(2) model with 
activity at alpha and gamma (30-50 Hz) bands.}
\end{figure}

\begin{figure}
\centering
\subfigure[ \label{F37c}]{\includegraphics[scale=.3]{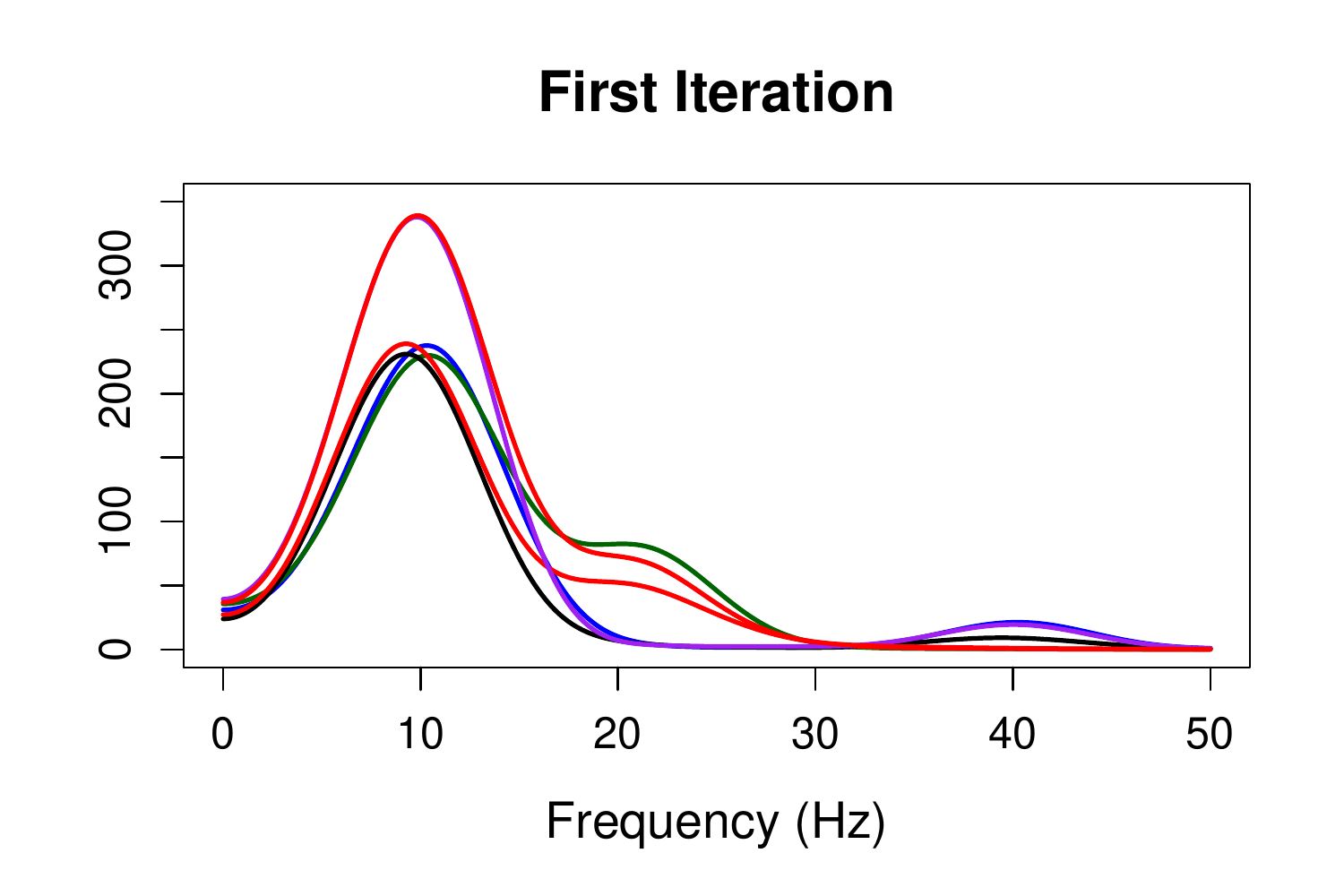}}
\subfigure[ \label{F37d}]{\includegraphics[scale=.3]{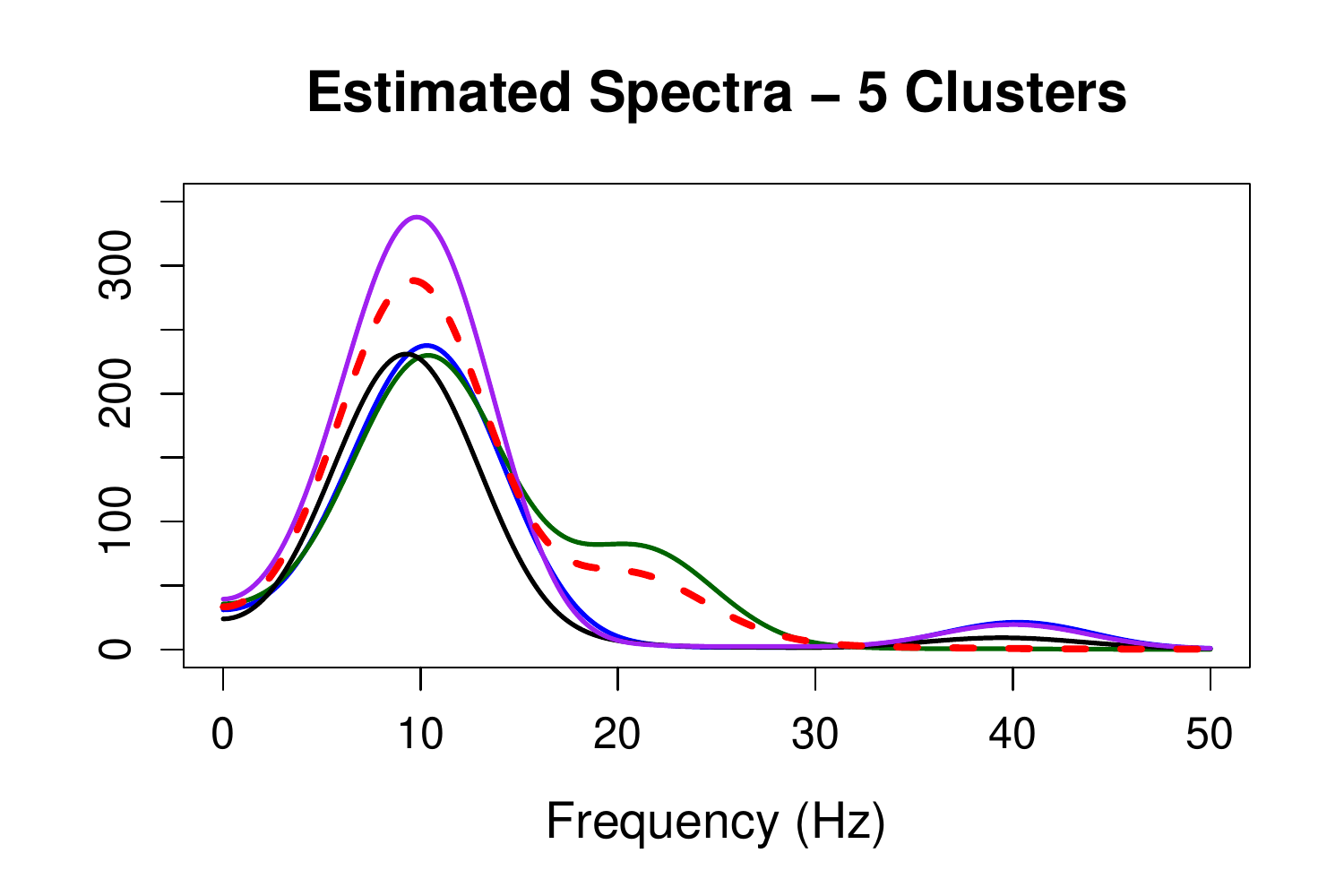}}
\subfigure[ \label{F37e}]{\includegraphics[scale=.3]{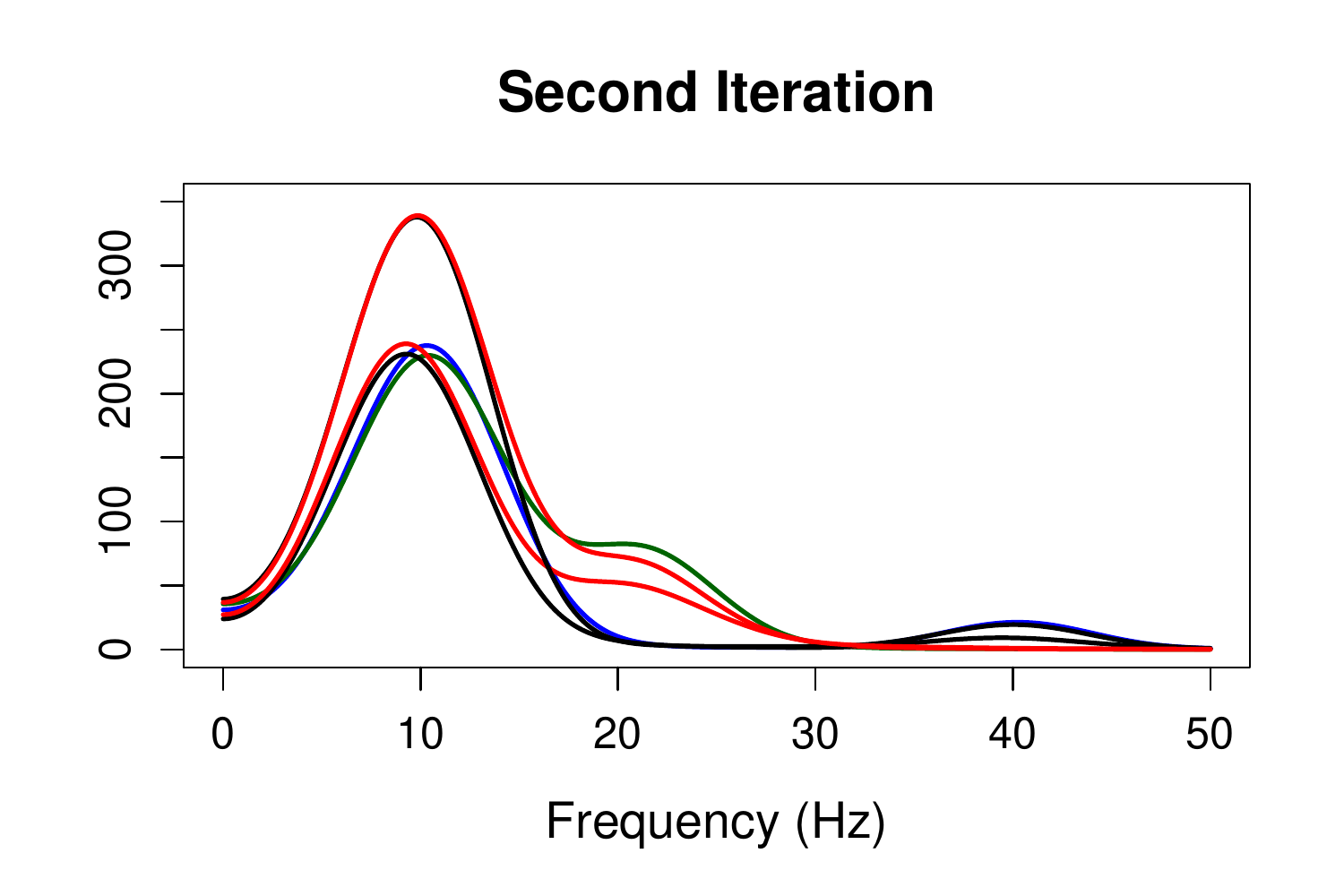}}
\subfigure[ \label{F37f}]{\includegraphics[scale=.3]{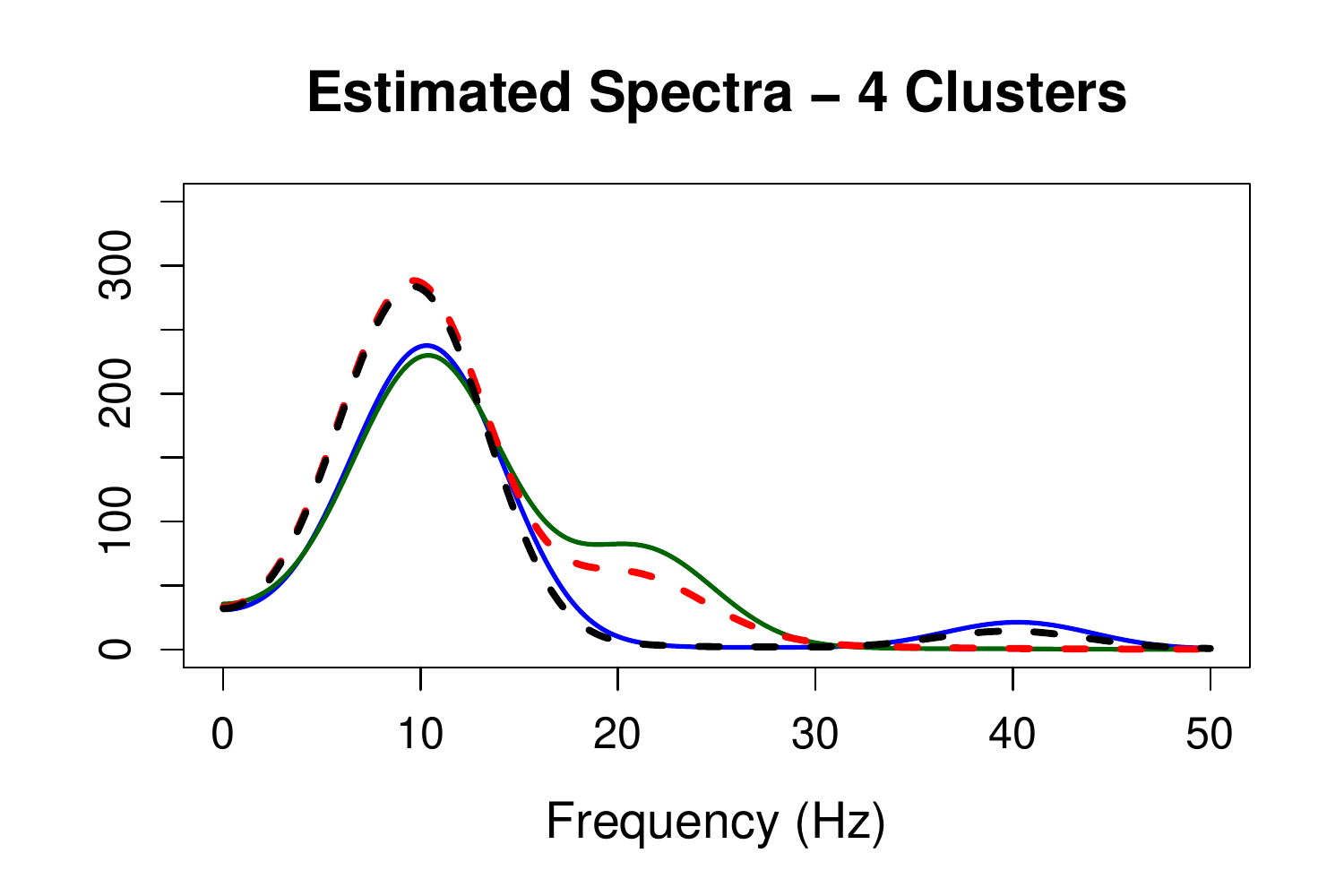}}
\subfigure[ \label{F37g}]{\includegraphics[scale=.3]{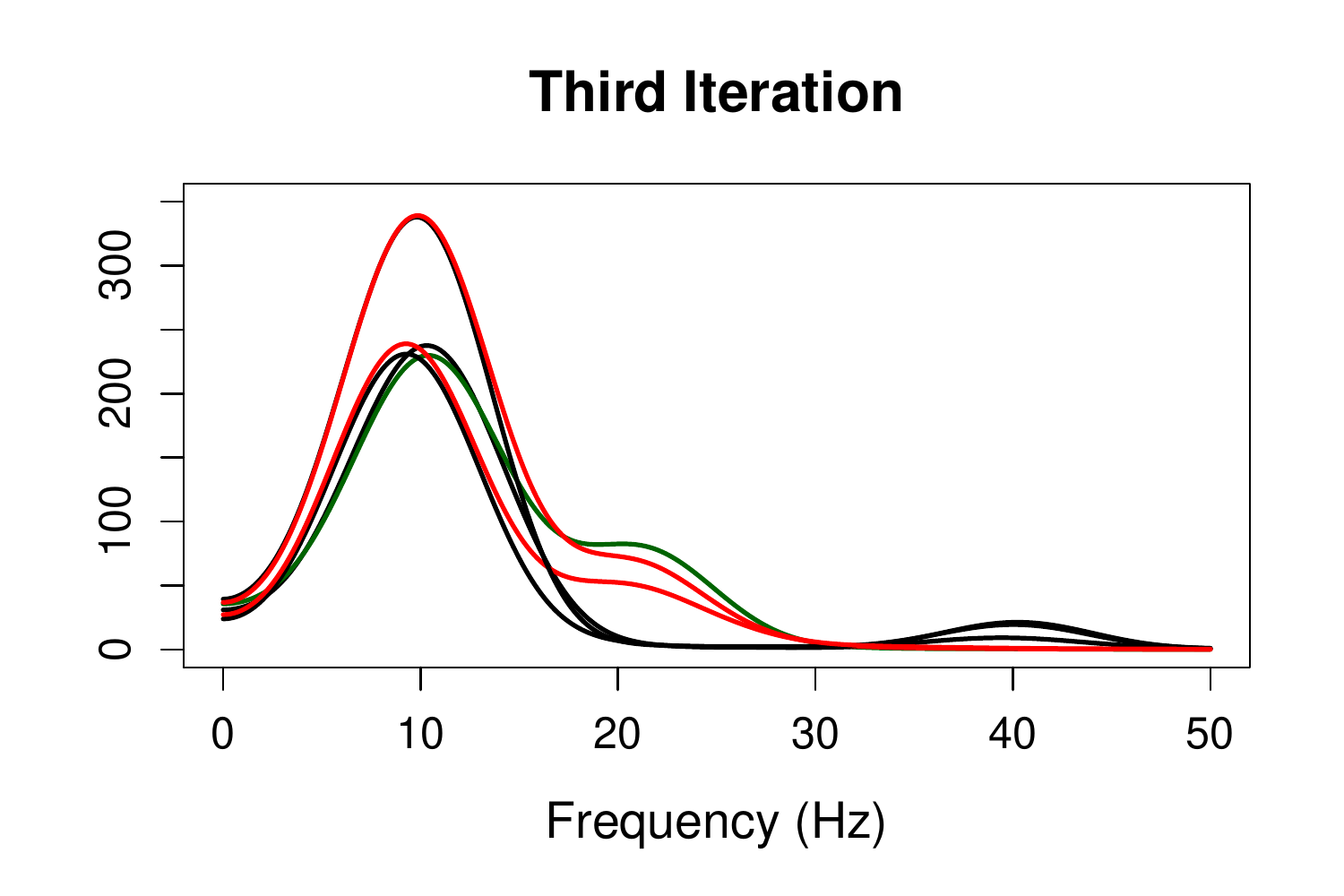}}
\subfigure[ \label{F37h}]{\includegraphics[scale=.3]{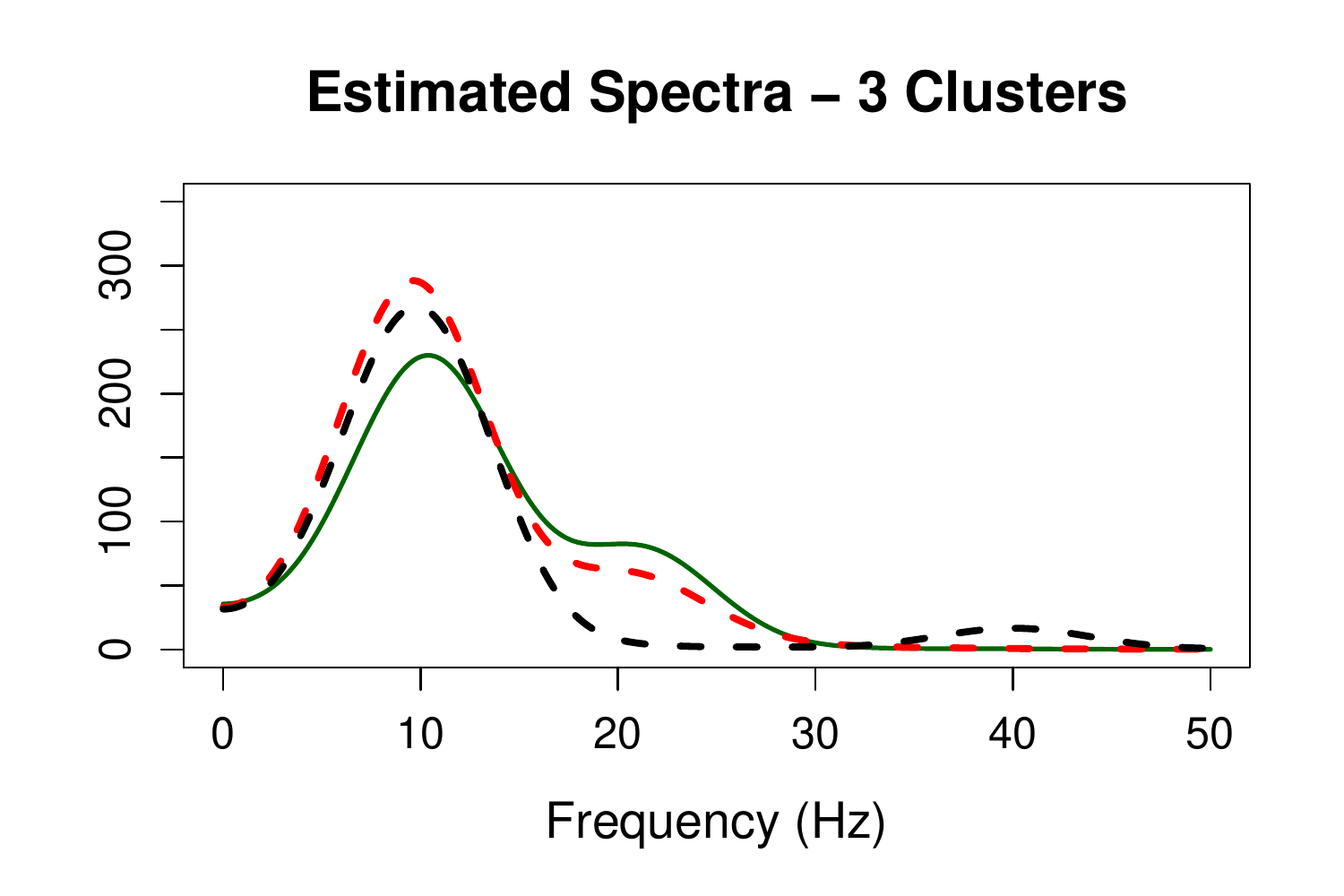}}
\subfigure[ \label{F37i}]{\includegraphics[scale=.3]{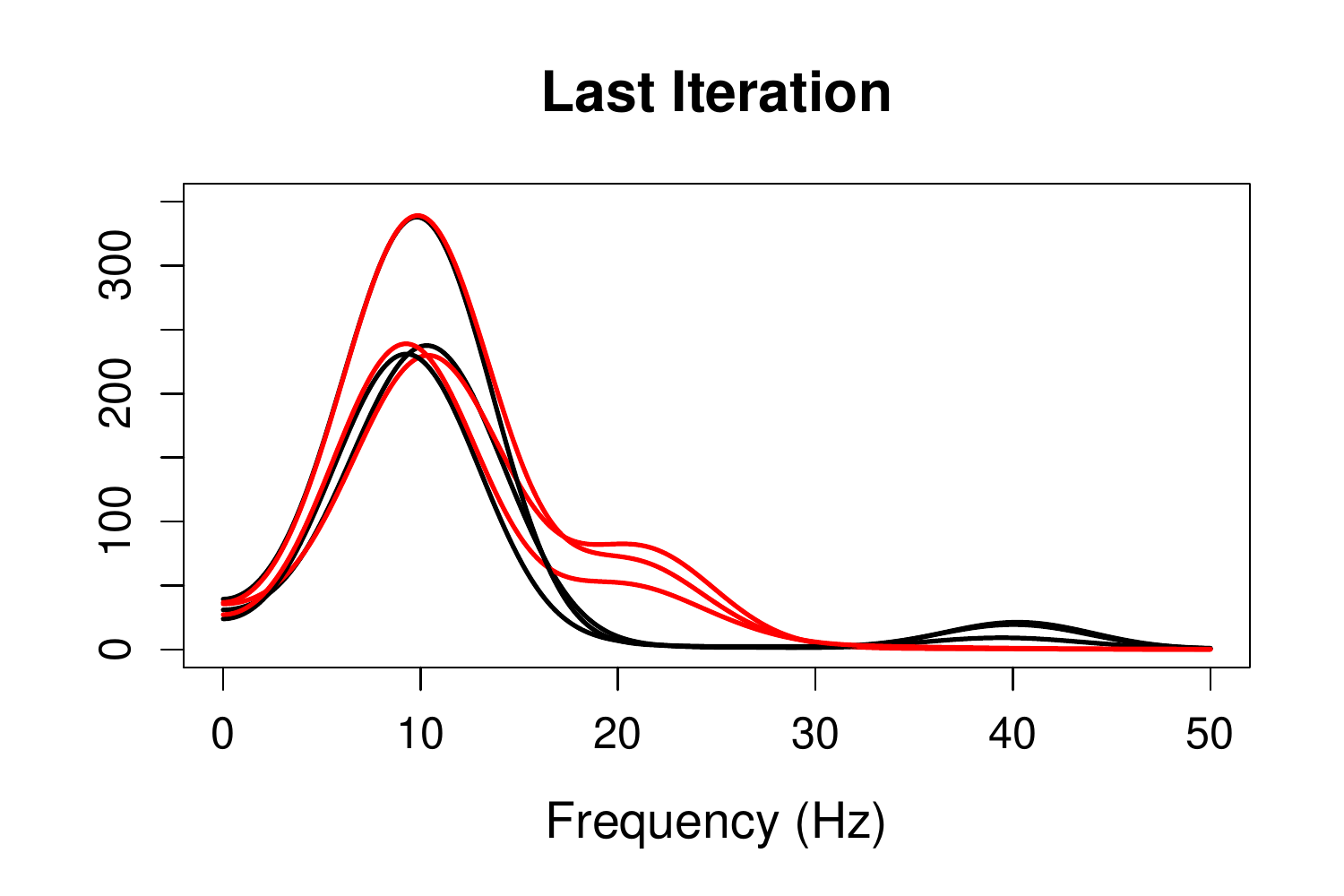}}
\subfigure[ \label{F37j}]{\includegraphics[scale=.3]{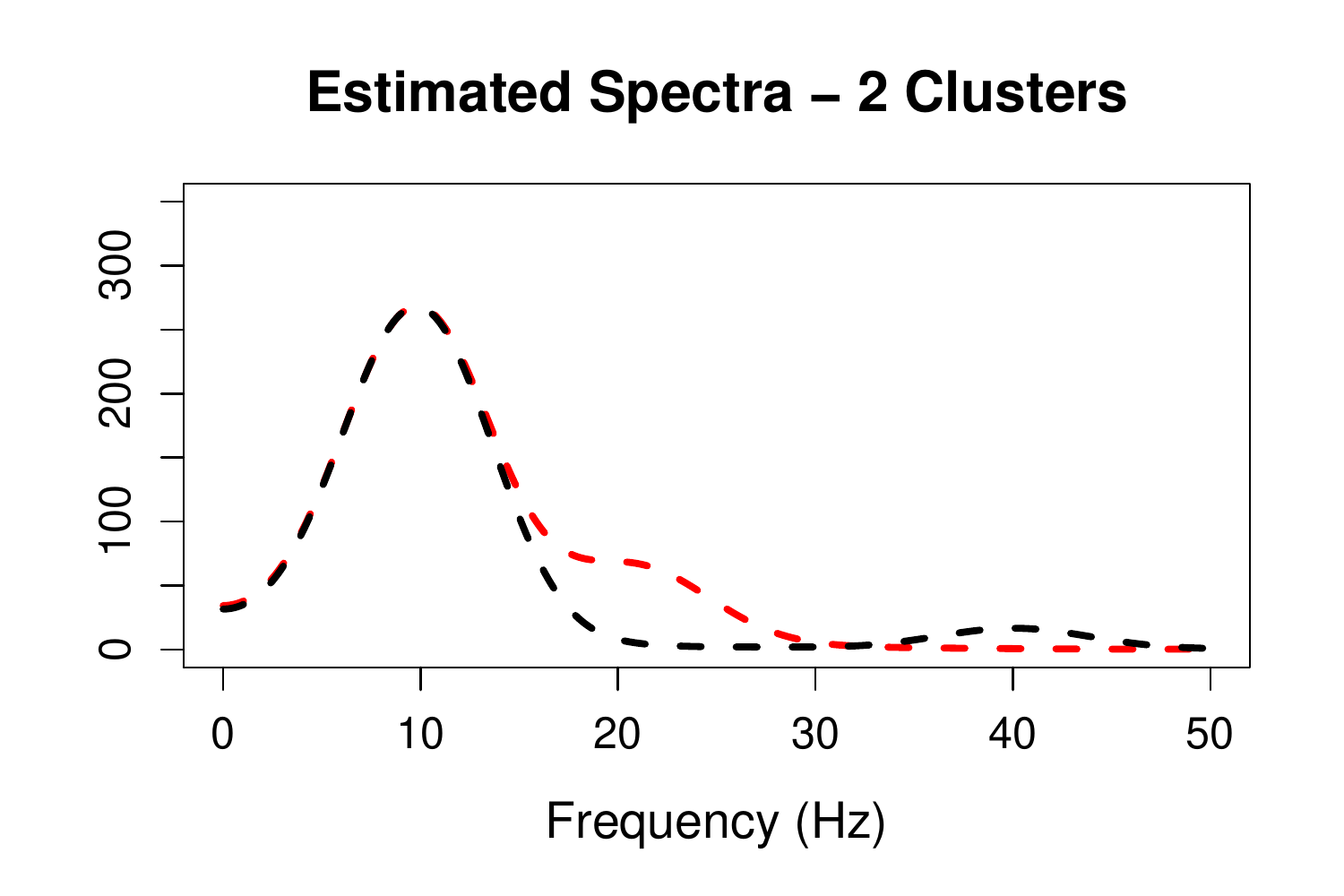}}
\caption{Dynamics of the hierarchical spectral merger algorithm. (a), (c), (e) 
and (g) 
show the clustering process for the spectra. (b), (d), (f) and (h) show the 
evolution of the estimated spectra, which improves as more series are merged in 
the same cluster.}\label{F37}
\end{figure}
\textit{\bf{Illustration.}}
Consider two different AR(2) models with their 
spectra concentrated at 10 Hertz, however, one also contains power at 21 Hertz 
while 
the other has power at 40 Hertz. We simulated three time series for each 
process, 
10 
seconds of each one with a sampling frequency of $100$ Hertz 
($t=1,\ldots,1000$). 
Figure \ref{F36a} shows the estimated spectra for each series and Figure 
\ref{F36b} shows by different colors (red and black) which one belongs to the 
first or second process. If we only look at the spectra, it is hard to 
recognize the number of clusters and their memberships. The method might have 
difficulty in clustering some cases, like the red and purple spectra.

The step-by-step result of the HSM method is shown in Figure \ref{F37}. We 
start 
with six 
clusters; at the first iteration we find the closest spectra, represented in 
Figure \ref{F37c} with the same color (red). After the first iteration we merge 
these time series and get $5$ estimated spectra, one per cluster, Figure 
\ref{F37d} shows the estimated spectra where the new cluster is represented by 
the dashed red curve. We can follow the procedure in Figures \ref{F37e}, 
\subref{F37f}, \subref{F37g} and \subref{F37h}. In the end, the proposed 
clustering algorithm reaches the correct solution, Figures \ref{F37i} and 
\ref{F36b} coincide. Also, the estimated spectra for the two clusters, Figure 
\ref{F37j}, is better than any of the initial spectra and we can identify the 
dominant frequency bands for each cluster.
\section{Numerical Experiments} 
We now investigate the finite sample performance of the HSM clustering 
algorithm.
First, we explain the simulation methods based on the spectrum that we will use 
in some of the experiments. Then, we present the results of the experiments, 
assuming 
that the number of clusters is known. Finally, we explore the case of  unknown 
number of clusters and possible criteria to choose it. 

\bigskip
\textit{Simulation based on a parametric family of spectral densities.}

\begin{figure}
 \centering
\includegraphics[scale=.35]{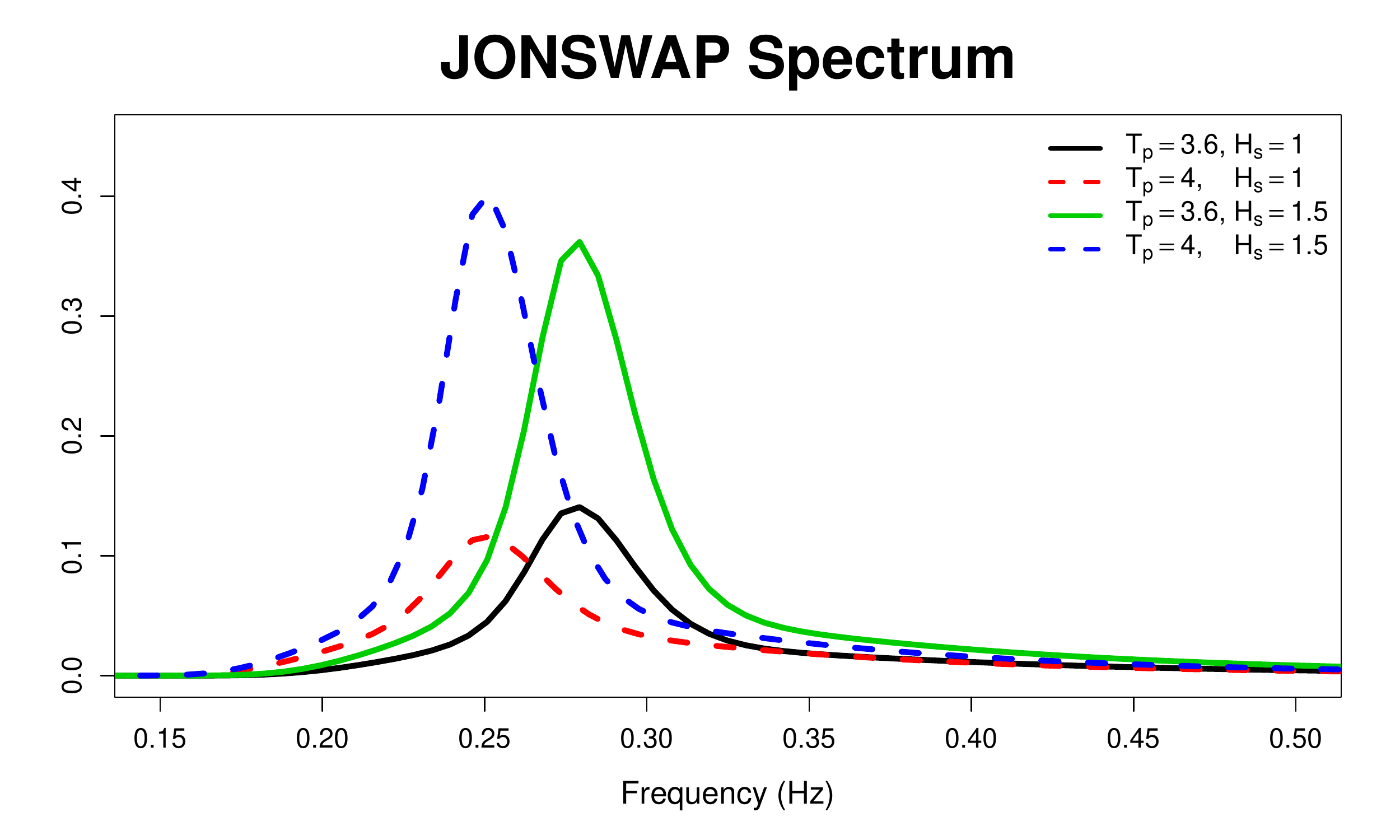}
\caption{JONSWAP spectrum with different parameters.}\label{JonsSpec}
\end{figure}

There exist several parametric families of spectral densities of frequent use  
in 
oceanography, which have an interpretation in terms of the behavior of sea 
waves (\cite{Ochi98}).
Motivated by the applications, we will simulate time series (Gaussian process) 
using spectra from one of these families. This methodology is already 
implemented by \cite{Wafo11} in the WAFO toolbox for MATLAB. 

An example of a group of parametric densities is the JONSWAP (Joint North-Sea 
Wave Project) spectral family, which is given by the formula
 $$
 S(\omega) = \frac{g^2}{\omega^5}\exp(-5\omega_p^4/4\omega^4)
 \gamma^{\exp(-(\omega-\omega_p)^2/2\omega_p^2s^2)}, \qquad \omega \in 
(-\pi,\pi),
 $$
where  $g$ is the acceleration of gravity, $s = 0.07$ if $\omega\leq
\omega_p$ and $s=0.09$ otherwise; $\omega_p=\pi/T_p$ and $\gamma =
\exp(3.484(1-0.1975 (0.036-0.0056 T_p/\sqrt{H_s}) T_p^4/(H_s^2)))$.
The parameters for the model are the significant wave height $H_s$,
which is defined as 4 times the standard deviation of the time series,
and the spectral peak period $T_p$, which is the period
corresponding to the modal frequency of the spectrum. Figure 
\ref{JonsSpec} shows some examples of this family with different values for the 
parameters $T_p$ and $H_s$. 
\vspace{.5cm}
\\
\textit{Simulation based on AR(2) processes.}
Consider the second order auto-regressive AR(2) model which is defined as
\begin{equation}
 Z_t=\phi_1 Z_{t-1}+ \phi_2 Z_{t-2}+ \epsilon_t,
\end{equation}
where $\epsilon_t$ is a white noise process. The characteristic polynomial for 
this model is $\phi(z)= 1-\phi_1 z-\phi_2 z^2$. The roots of the polynomial 
equation (assumed with magnitude bigger than $1$ for causality) indicate the 
properties of the
oscillations. If the roots, denoted $z_0^{1}$ and $z_0^{2}$ are complex-valued 
then they must be complex-conjugates, i.e., $z_0^{1}=\overline{z_0^{2}}$. These 
roots have a polar representation
\begin{equation}\label{AR2}
 |z_0^{1}|=|z_0^{2}|=M, \qquad \qquad \arg(z_0)= \frac{2 \pi \eta}{F_s},
\end{equation}
where $F_s$ denotes the sampling frequency (in Hertz); $M$ is the amplitude or 
magnitude of the root
($M >1$ for causality); and $\eta$ is the frequency index ($\eta \in 
(0,F_s/2)$). The spectrum of the 
AR$(2)$ process
with polynomial roots as above will have peak frequency at $\eta$. The peak 
becomes 
broader as $M \to \infty$, and it becomes narrower as $M \to 1^{+}$.

\begin{figure}
\centering
\includegraphics[scale=.3]{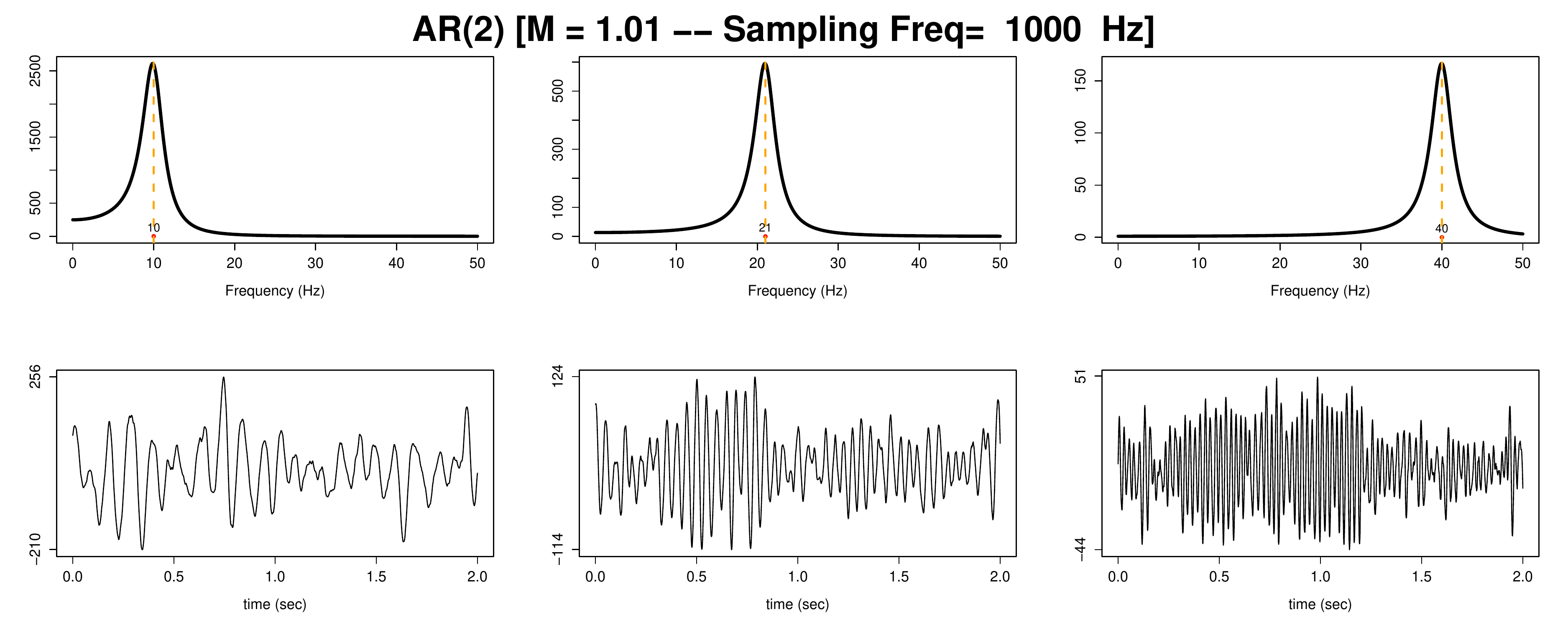}
\caption{Top: Spectra for the AR(2) process with different peak frequency; 
$\eta=10,21,40$.
Bottom: Realizations from the corresponding AR(2) process.}\label{F2}
\end{figure}

Then, given $(\eta,M,F_s)$, we take  
\begin{align}\label{AR2_2}
  \phi_1=\frac{2 \cos(w_0)}{M}\qquad \mbox{and}\qquad
  \phi_2=-\left(\frac{1}{M^2}\right), 
\end{align}
where $w_0=\frac{2\pi\eta}{F_s}$. If one computes the roots of the 
characteristic polynomial with the coefficients in \eqref{AR2_2}, they satisfy 
\eqref{AR2}. 

To illustrate the type of oscillatory patterns that can be observed in time 
series
from processes with corresponding spectra,  we plot in Figure \ref{F2} the 
spectral densities (top)
for different values of $\eta$, $M=1.01$ and $F_s=1000$ Hertz; and examples of 
generated time series (bottom).
Larger values of $\eta$ gives rise to faster oscillations (higher 
frequencies) of the signal.

\subsection{Experimental Design}
\begin{figure}
\centering
\subfigure[\label{SE1}Experiment1]
{\includegraphics[scale=.4]{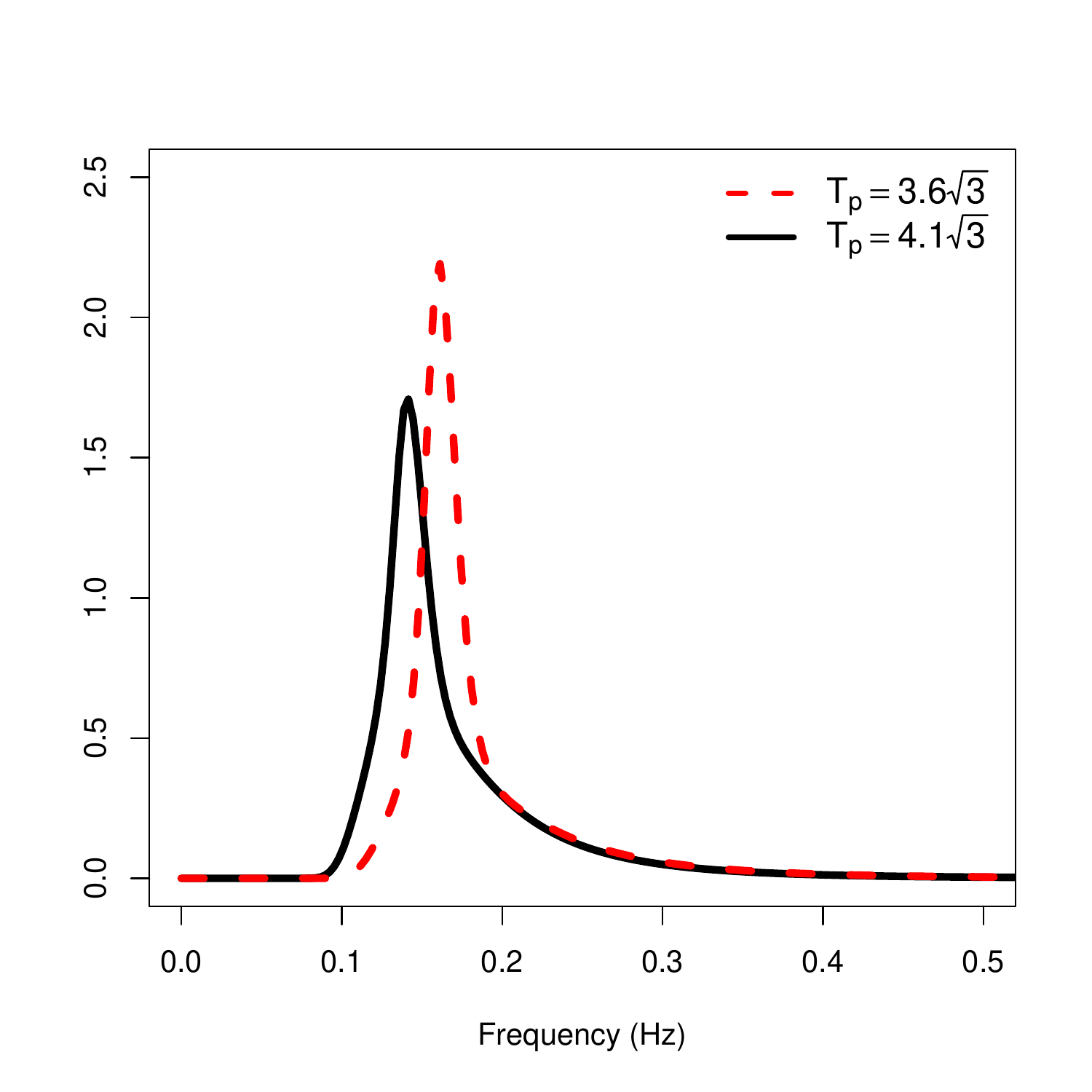}}\hspace{1cm}
\subfigure[\label{SE2}Experiment 2]
{\includegraphics[scale=.4]{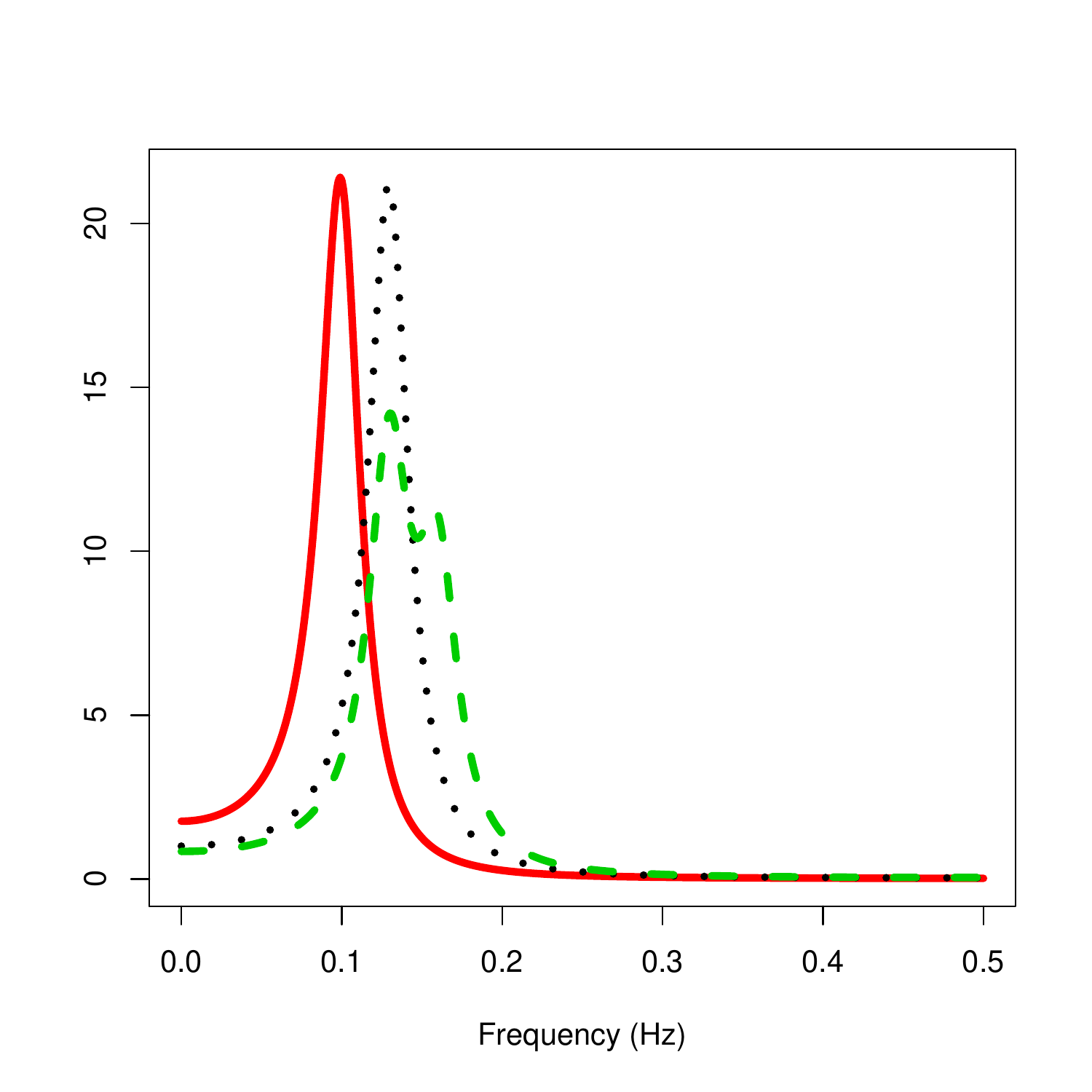}}
\caption{Spectra used in the simulation study to compare the HSM method with 
other similarity measures. Each spectrum, with different color and line type, 
corresponds to a cluster.}\label{SpecExp}
\end{figure}

We consider two different experiments. The first one is motivated by  
applications in oceanography, where the differences between spectra could be 
produced by a small change in the modal frequency. The second experiment was 
designed to test if the algorithms are able to distinguish between unimodal 
and bimodal spectra. For all the experiments, the lengths of the time series 
were $T=500,1000,$ and $2000$. 
\begin{itemize}
 \item \textbf{Experiment 1} is based on two different JONSWAP (Joint North-Sea 
Wave Project) spectra (i.e. two clusters). The spectral densities considered 
have significant wave height 
$H_s$ 
equal to $3$, the first has a peak period $T_p$ of $3.6\sqrt{H_s}$ while for 
the 
second $T_p= 4.1\sqrt{H_s}$. Figure \ref{SE1} exhibits the JONSWAP spectra, 
showing that the curves are close to each other. Five series from each spectrum 
were simulated and $N=500$ replicates of this experiment were made. In this 
case the sampling frequency was set to 1.28 Hertz, which is a common value for 
wave data recorded using sea buoys. This experiment was carried out in 
\cite{Alvarez15}  to compare several clustering procedures.

 \item \textbf{Experiment 2} is based on the AR$(2)$ process. Let $Z_t^j$, 
$j=1,2,3$, be AR(2) processes with $M_j=1.1$ for all $j$ and 
peak frequency $\eta_j =.1,.13,.16$ for $j=1,2,3$, respectively. $Z_t^{j}$ 
represents a latent signal oscillating at a pre-defined band. Define the 
observed time series to be a mixture of these latent AR$(2)$ processes.
\begin{equation}\label{Sim1}
\begin{pmatrix}
 X_t^1\\
 X_t^2\\
 \vdots\\
 X_t^K\\
\end{pmatrix}_{K \times 1}
=
\begin{pmatrix}
 \boldsymbol{e}_1^T\\
 \boldsymbol{e}_2^T\\
 \vdots\\
 \boldsymbol{e}_K^T\\
\end{pmatrix}_{K \times 3}
\begin{pmatrix}
 Z_t^1 \\
 Z_t^2 \\
 Z_t^3\\
\end{pmatrix}_{3 \times 1}
+
\begin{pmatrix}
 \varepsilon_t^1\\
 \varepsilon_t^2\\
 \vdots\\
 \varepsilon_t^K\\
\end{pmatrix}_{K \times 1}
\end{equation}
where $\varepsilon_t^j$ is Gaussian white noise, $X_t^j$ is a signal with 
oscillatory behavior generated by the linear combination $\boldsymbol{e}_i^T 
{\bf Z}_t$ and $K$ is the number of clusters. In this experiment, we set $K=3$, 
with 
$\boldsymbol{e}_1^T=c(1,0,0),~ \boldsymbol{e}_2^T=c(0,1,0) $ and 
$\boldsymbol{e}_3^T=c(0,1,1)$. We simulate five replicates of each signal 
$X_t^{i}$, so, we have three clusters with five members each. Figure \ref{SE2} 
plots the three different spectra. For this experiment $N=1000$ replicates were 
made, and the sampling frequency was set to 1 Hertz.
\end{itemize}

\subsection{Comparative Study}\label{CS}
To compare clustering results, we must take into account the ``quality of the 
clustering" produced which depends on both the similarity measure and the 
clustering algorithm used. 
The HSM method has two main features: The use of the TV distance as a 
similarity 
measure and the hierarchical spectral merger algorithm.  
The HSM method will be compared with the usual hierarchical agglomerative 
clustering algorithm using the complete linkage function, which is one of the 
standard clustering procedures used in the literature. 

\cite{Vilar1} proposed two simulation tests to compare the
performance of several  dissimilarity measures for time series clustering. We 
compare the HSM method with competitors that were based on the spectral density 
and performed well in P\'ertega D\'iaz and Vilar's experiments. In addition, we 
also considered the 
distance based on the cepstral coefficients, which was
used in \cite{Maharaj11}, and the symmetric version of the Kulback-Leibler 
divergence, that was used in a hierarchical clustering algorithm in 
\cite{ShumStof}.

Let $ I_X(\omega_k) = T^{-1} \Big|\sum_{t=1}^T X_t e^{-i\omega_k
t}\Big|^2 $ be the periodogram for time series $X$, at frequencies
$\omega_k=2\pi k/T, \ k=1, \dots, n$ with $n=[(T-1)/2]$, and $NI_X$
be the normalized periodogram, i.e. $NI_X(\omega_k) =
I_X(\omega_k)/\hat\gamma_0^X $, with $\hat\gamma_0^X$ 
the sample variance of time series $X$ (notice that $\hat\gamma_0^X= 
\sum_{k=-(n-1)}^n I_X(\omega_k)$). The estimator of the 
spectral density $ \widehat{f}_{X}$ is 
the smoothed periodogram using a Parzen window with bandwidth equal to $100/T$, 
normalized by dividing by $\hat\gamma_0^X$.

The dissimilarity criteria in the
frequency domain considered were:
\begin{itemize}
\item The Euclidean distance between the normalized estimated spectra: 
$
d_{NP}(X,Y) = \frac{1}{n}\Big( \sum_k \big( \widehat{f}_{X}(\omega_k)- 
\widehat{f}_{Y}(\omega_k)\big)^2\Big)^{1/2}.
$
\item The Euclidean distance between the logarithm of the normalized estimated 
spectra: 
$
d_{LNP}(X,Y) = \frac{1}{n}\Big( \sum_k \big( \log \widehat{f}_{X}(\omega_k)- 
\log \widehat{f}_{Y}(\omega_k)\big)^2\Big)^{1/2}.
$

\item The square of the Euclidean distance  between the cepstral coefficients
$
d_{CEP}(X,Y) =  \sum_k^p \big( \theta_k^X-\theta_k^Y\big)^2
$
where, $\theta_0=\int_0^1 \log I(\lambda) \mbox{d}\lambda$ and
$\theta_k=\int_0^1 \log I(\lambda) \cos(2 \pi k \lambda )
\mbox{d}\lambda$.

\item The symmetric Kullback-Leibler distance between the normalized estimated 
spectra: 
$
d_{KL}(X,Y)= \int \widehat{f}_{X} (\omega) 
\log\left(\frac{\widehat{f}_X(\omega)}{\widehat{f}_Y(\omega)}\right) 
\mbox{d}\omega, \quad
d_{SKL}(X,Y)=d_{KL}(X,Y)+d_{KL}(Y,X).
$
\end{itemize}

We also added the clustering method proposed by \cite{Alvarez15}, that uses the 
TV distance in a hierarchical clustering algorithm. All these dissimilarity 
measures were compared with the HSM method using 
normalized estimated spectra.

To evaluate the rate of success, we considered the following index which has 
been 
already used for comparing different clustering procedures [\cite{Vilar1}, 
\cite{Gav00}]. Let $\{C_1,\ldots,C_g\}$ and $\{G_1,\ldots,G_k\}$, be the set 
of the $g$ true groups 
and a $k$-cluster solution, respectively.
Then, $ \displaystyle \mbox{Sim}(G,C)=\frac{1}{g}\sum_{i=1}^{g} \max_{1\leq 
j\leq k} \mbox{Sim}(G_j,C_i),$
where $ \displaystyle \mbox{Sim}(G_j,C_i)=\frac{2|G_j \cup C_i|}{|G_j|+|C_i|}$. 
Note that this similarity measure will return 0 if the two clusterings are 
completely dissimilar and 1 if they are the same.

\begin{figure}
\centering
\subfigure[$T=500$.]
{\includegraphics[scale=.26]{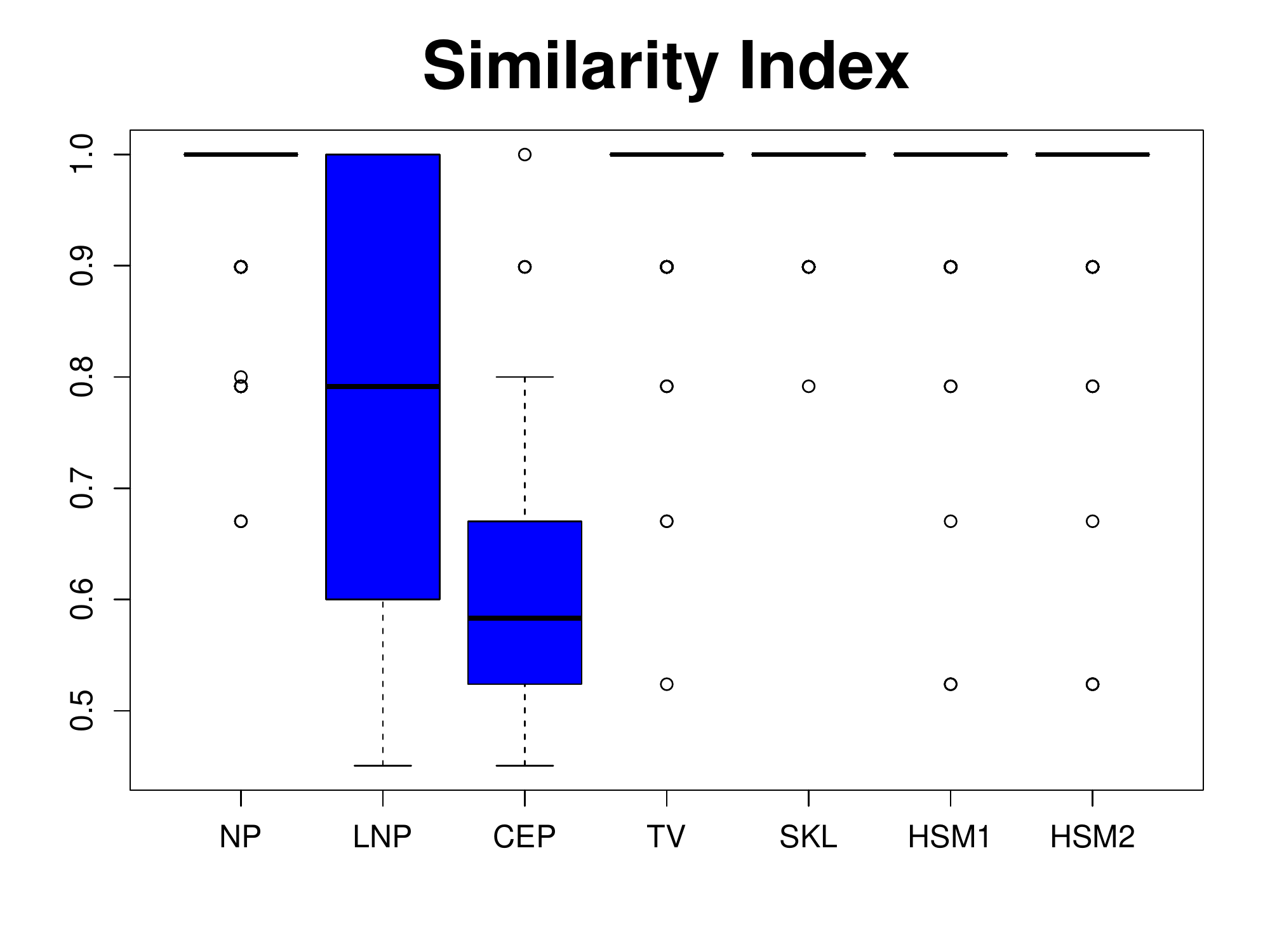}}
\subfigure[$T=1000$.]
{\includegraphics[scale=.26]{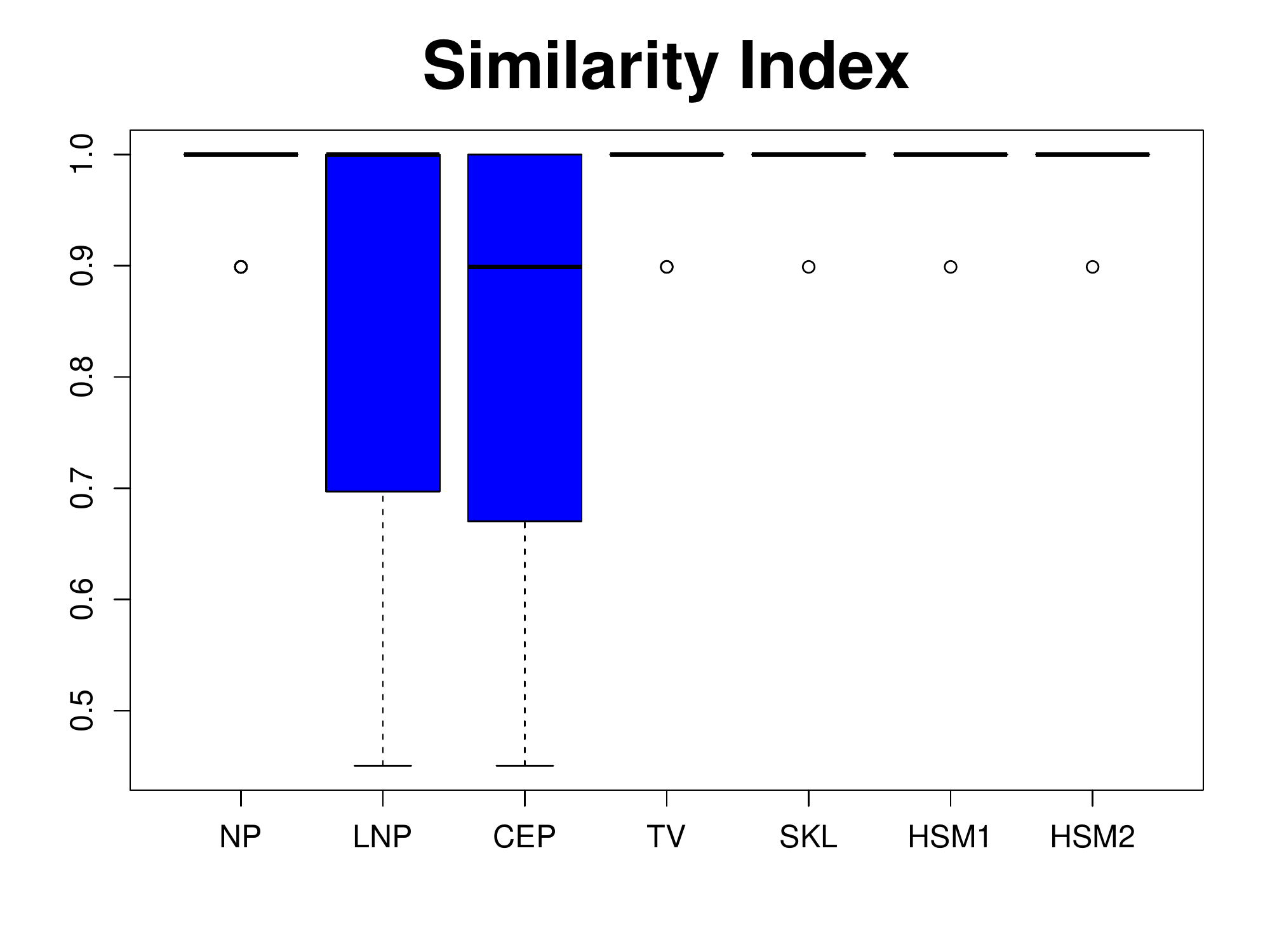}}
\subfigure[$T=2000$.] 
{\includegraphics[scale=.26]{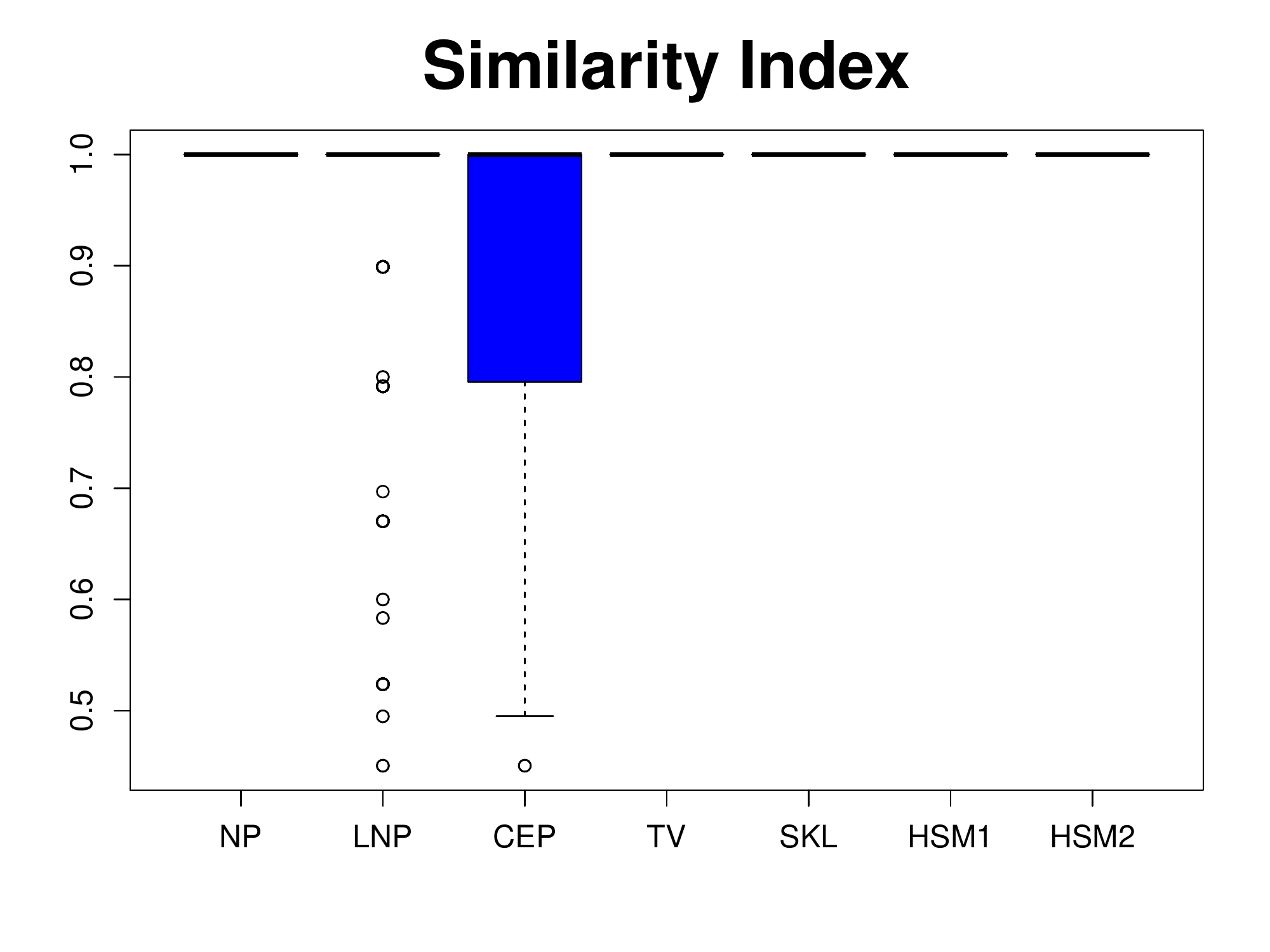}}
\caption{Boxplots of the rate of success for the replicates under the 
simulation setting of Experiment 1, by using different 
distances.}\label{BPExp1Jons}
\end{figure}

\begin{figure}
\centering
\subfigure[$T=500$.]
{\includegraphics[scale=.26]{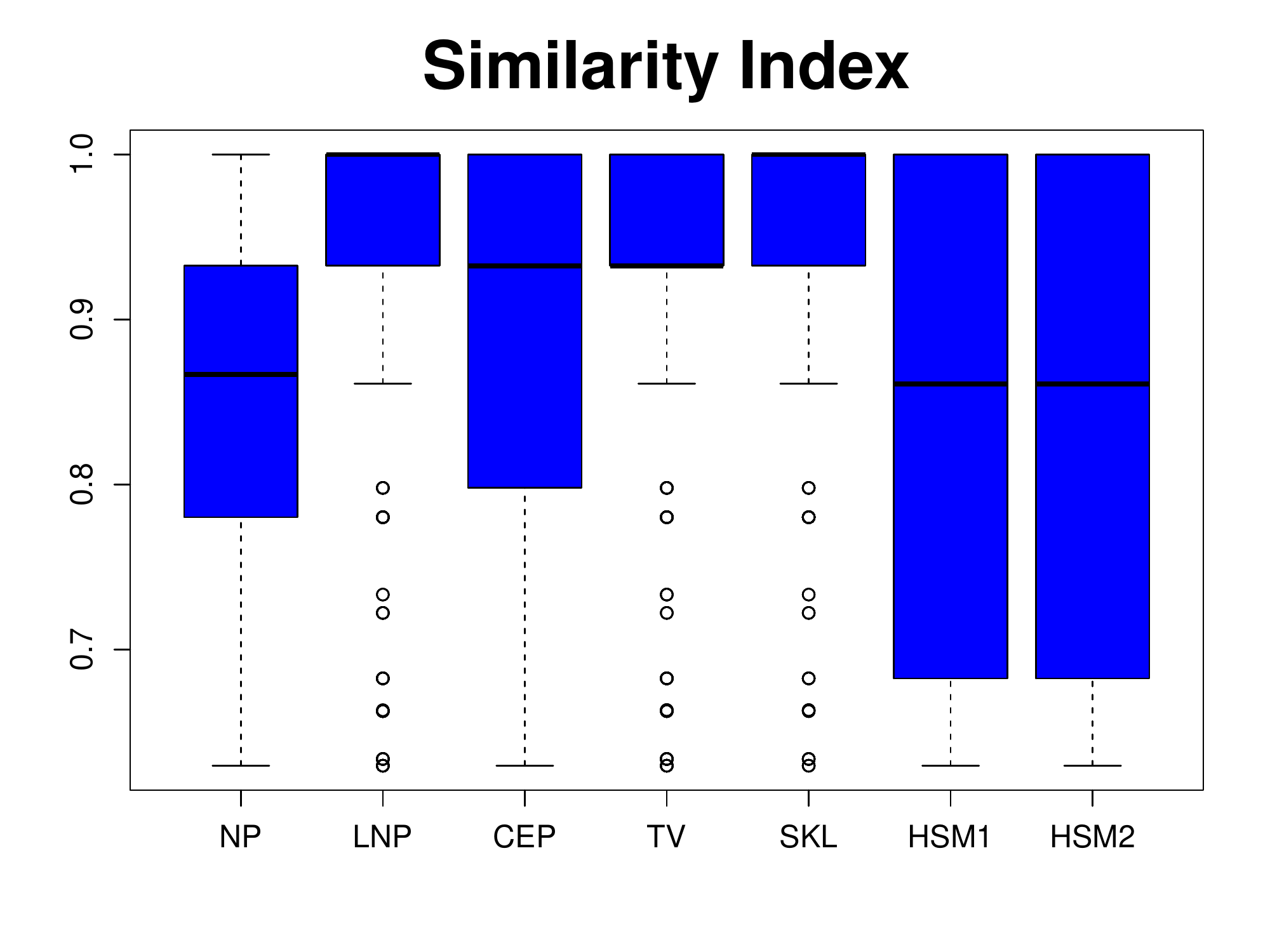}}
\subfigure[$T=1000$.]
{\includegraphics[scale=.26]{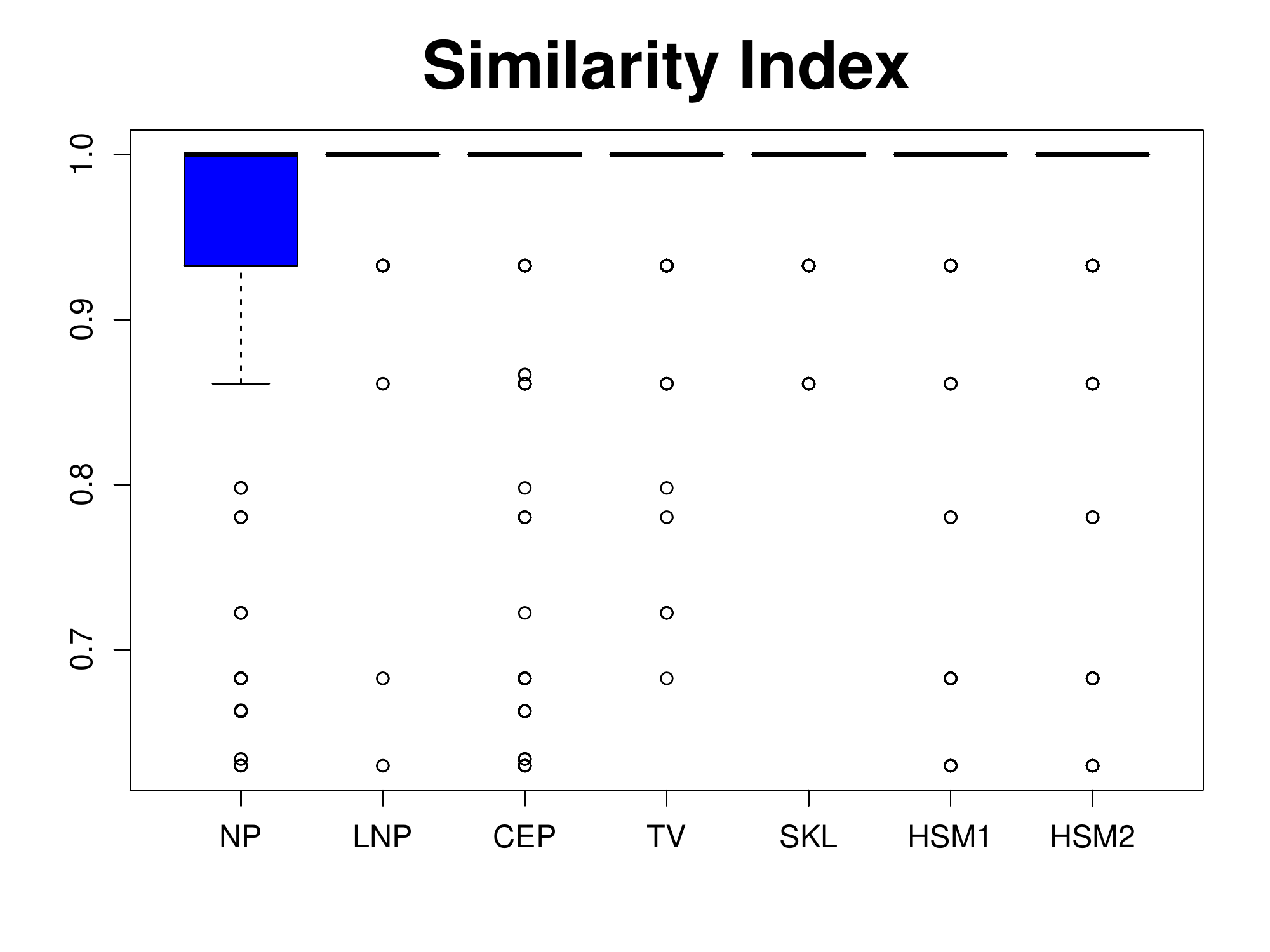}}
\subfigure[$T=2000$.] 
{\includegraphics[scale=.26]{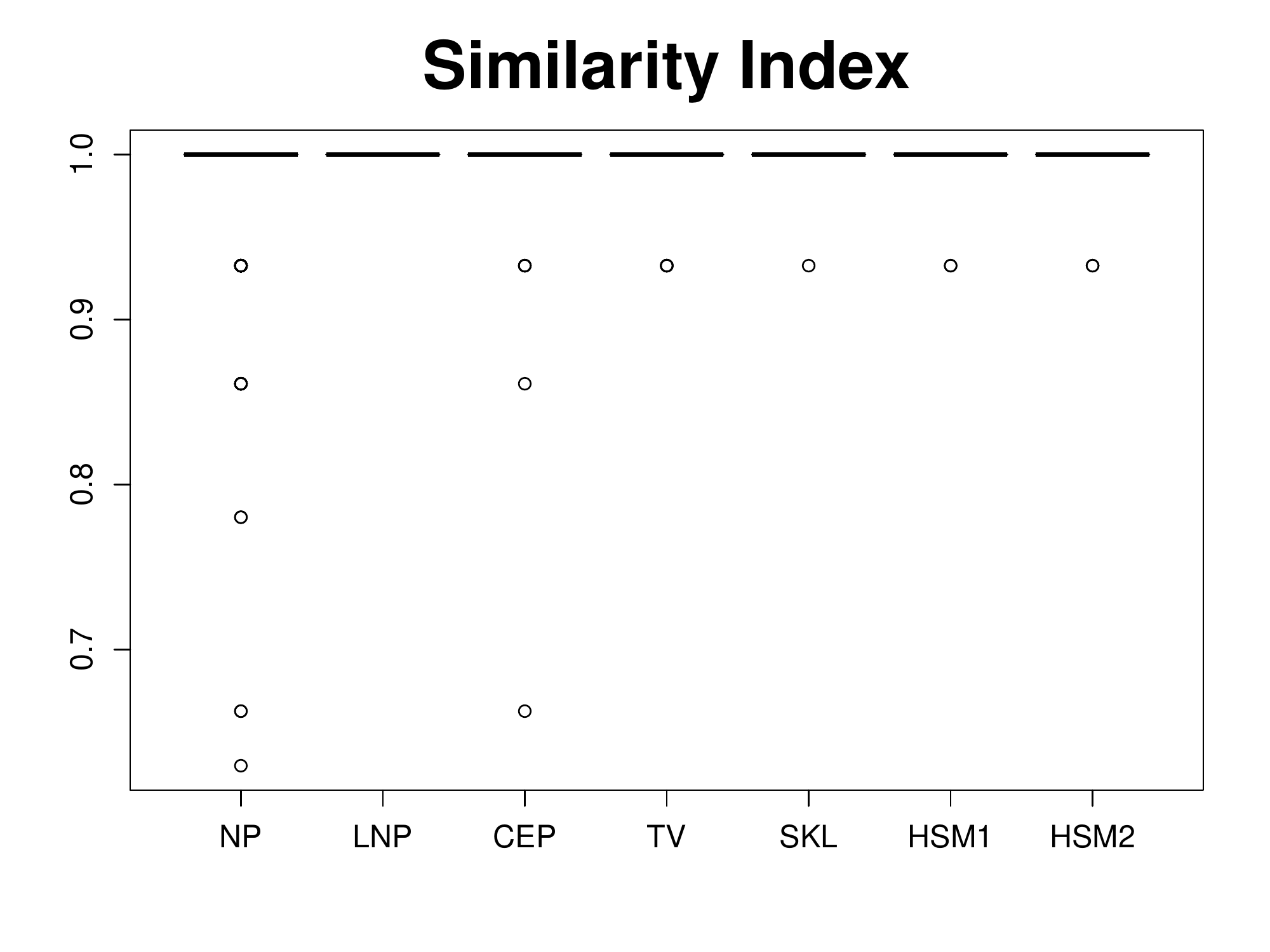}}
\caption{Boxplots of the rate of success for the replicates under the 
simulation setting of Experiment 2, by using different 
distances}\label{BPExp2AR}
\end{figure}

In the comparative study, each simulation setting was replicated N times, and 
 the rate of success for each one was computed. The mean values for this index 
are shown in 
Tables \ref{Exp1Jons} and \ref{Exp2AR2}, and  boxplots of the 
values obtained are shown in Figures \ref{BPExp1Jons} and \ref{BPExp2AR}. The 
simulation settings were the same in all cases  and the clustering algorithm is 
hierarchical with the complete link 
function (similar results are obtained with the average link). For the HSM 
method, we used the notation HSM1 when we used the single version, and 
HSM2 when we used the average version. The 
clustering method of \cite{Alvarez15} is denoted by TV.     
\begin{table}\footnotesize
\centering
\begin{tabular}{cccccccc}
Experiment 1 & \\
\hline \hline\\
$T$  & NP&	LNP&	CEP&	TV&	SKL&	HSM1&	HSM2\\
\hline \\
500&	         0.979&	0.772&	0.597&	0.988&	0.994&	0.989&	0.988\\
1000&	0.998&	0.851&	0.825&	0.999&	0.999	&0.999&	0.999\\
2000&	1&	0.932&	0.908&	1&	1& 1&	1\\
\end{tabular}
\caption{Mean values of the similarity index obtained using different distances 
and the two proposed methods in Experiment 1. The number of replicates is 
$N=500$.}\label{Exp1Jons}
\end{table}
\begin{table}\scriptsize
\centering
\begin{tabular}{cccccccc}
Experiment 2 & \\
\hline \hline\\
$T$  & NP&	LNP&	CEP&	TV&	SKL&	HSM1&	HSM2\\
\hline \\
500&		0.864&	0.949&	0.895&	0.930&	0.952&	0.836&	0.838\\
1000&	0.961&	0.996&	0.974&	0.990&	0.994&	0.983&	0.983\\
2000&	0.995&	1&	0.999&	0.999&	0.999&	0.999&	0.999\\
\\
\end{tabular}
\caption{Mean values of the similarity index obtained using different distances 
and the two proposed methods in Experiment 2. The number of replicates is 
$N=1000$.}\label{Exp2AR2}
\end{table}

We can see from the boxplots corresponding to \textbf{Experiment 1} that the 
CEP distance has many values smaller than $0.9$ even in the case of $T=2000$.
In \textbf{Experiment 2}, the HSM method did not have a good performance 
for small-sized time series. It is necessary to have $T=1000$, for the HSM 
method to identify the clusters more precisely. The 
NP distance has the worst performance overall. 

In general, the rates of success for the HSM methods are very close to one. In 
some cases the HSM method has the best results, and when it is 
not the case, the rates obtained by the HSM method are close to the best. The 
methods based on logarithms, such as 
the LNP and CEP, have in some cases a good performance but in others their 
performance is very poor. Compared to the symmetric Kullback-Leibler distance, 
there is no clear winner but the method 
proposed here still has the advantage of
being easily interpretable because the KL (or symmetric KL) cannot indicate if 
the dissimilarity value
is large since it belongs to the range $[0, \infty)$.

\subsection{When the number of clusters is unknown}

For real data the number of clusters is usually unknown. Thus, an objective 
criterion is
needed to determine the optimal number of clusters. As mentioned in Step 3 of 
our
algorithm, the TV distance computed before joining two clusters can be used as 
a 
criterion.

We present two options for selecting the number of clusters. The first is 
an empirical criterion while the second one is based on a bootstrap procedure.

\subsubsection{Empirical criterion}
Consider the following experimental design, similar to that of 
\textbf{Experiment 2}. Let $Z_t^j$ be an AR(2) process 
with $M_j=1.01$ for all $j$ and peak frequencies $\eta_j = 2, 6, 10, 21$ and 
$40$ 
for $j=1, \ldots, 5$, respectively. 
Define the observed time series to be a mixture of these latent AR$(2)$ 
processes.
\begin{equation}\label{Sim1a}
\begin{pmatrix}
 X_t^1\\
 X_t^2\\
 \vdots\\
 X_t^K\\
\end{pmatrix}_{K \times 1}
=
\begin{pmatrix}
 \boldsymbol{e}_1^T\\
 \boldsymbol{e}_2^T\\
 \vdots\\
 \boldsymbol{e}_K^T\\
\end{pmatrix}_{K \times 5}
\begin{pmatrix}
 Z_t^1 \\
 Z_t^2 \\
 \vdots \\
 Z_t^5\\
\end{pmatrix}_{5 \times 1}
+
\begin{pmatrix}
 \varepsilon_t^1\\
 \varepsilon_t^2\\
 \vdots\\
 \varepsilon_t^K\\
\end{pmatrix}_{K \times 1}
\end{equation}
where $\varepsilon_t^j$ is Gaussian white noise, $X_t^j$ is a signal with 
oscillatory behavior generated by the linear combination $\boldsymbol{e}_i^T 
{\bf Z}_t$, $K$ is the number of spectrally synchronized groups or clusters 
and $n$ denotes the number of replicates of each signal $X_t^{i}$.

\begin{figure}
\centering
\subfigure[\label{F6a}]{\includegraphics[scale=.2]
{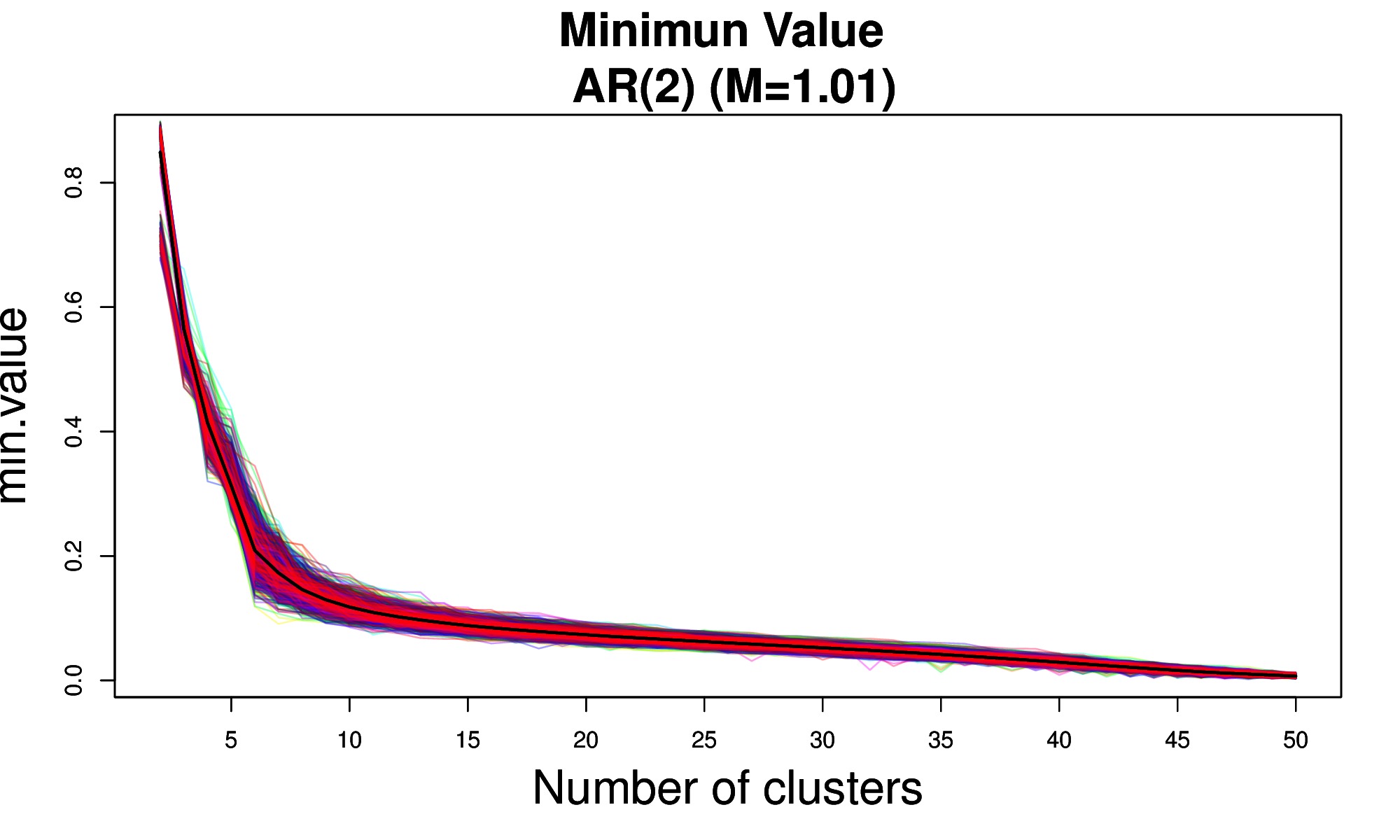}}
\subfigure[\label{F6b}]{\includegraphics[scale=.2]
{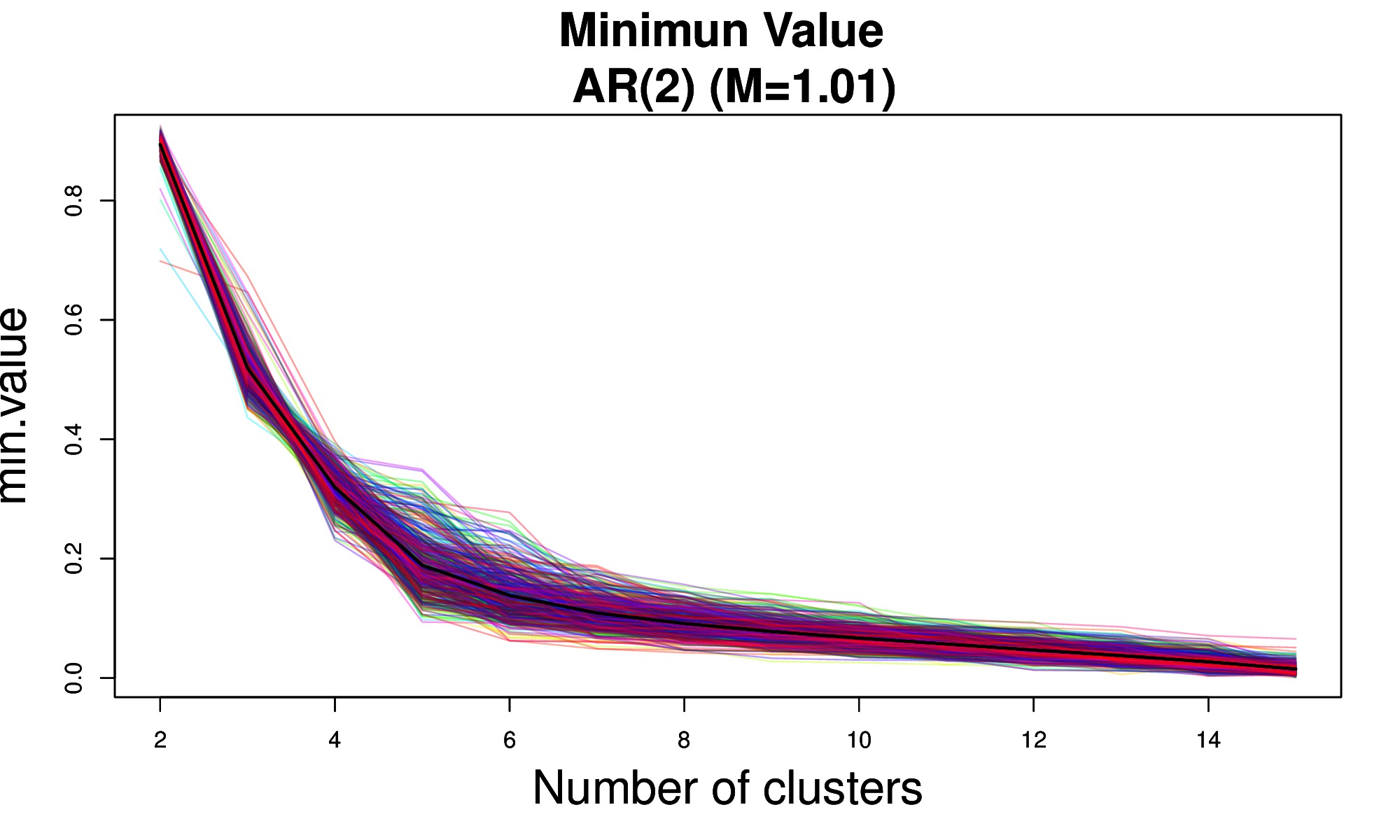}}
\subfigure[\label{F6c}]{\includegraphics[scale=.2]
{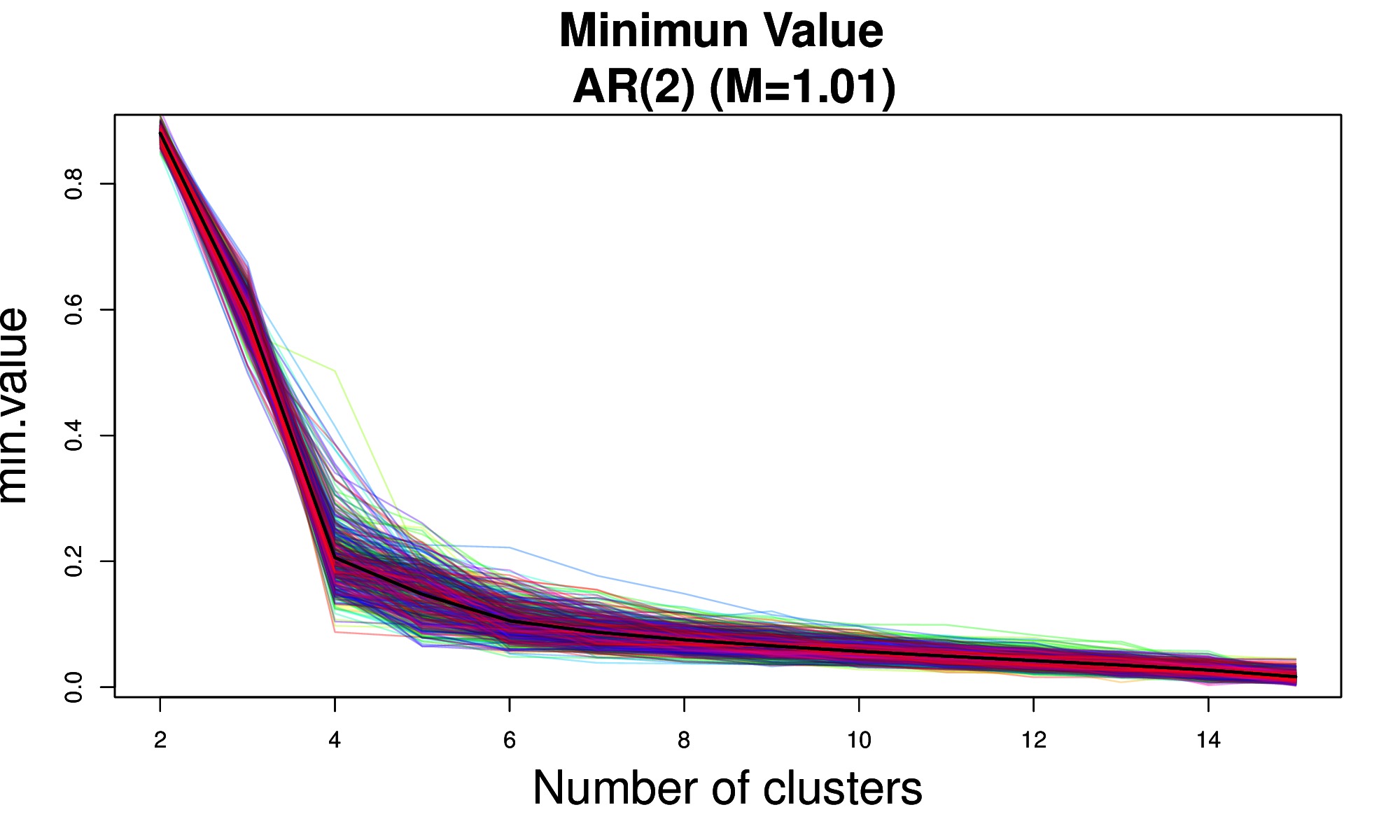}}
\subfigure[\label{F6d}]{\includegraphics[scale=.2]
{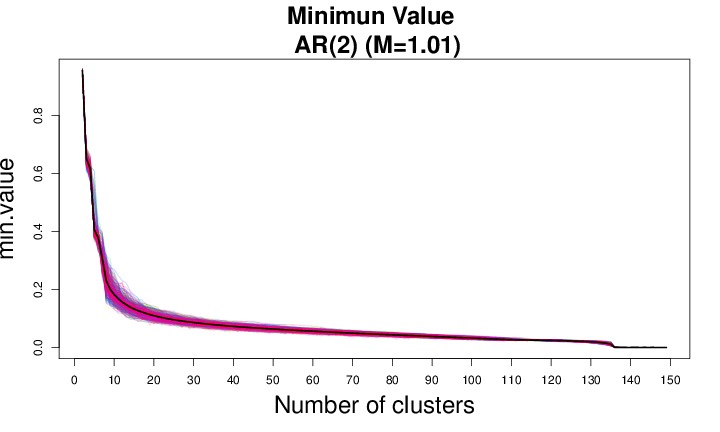}}
\caption{Trajectory of the value of the TV distance that is achieved by the 
algorithm.}\label{F6}
\end{figure}

Figure \ref{F6} displays the graphs corresponding to the minimum values 
of the TV distance between clusters as a function of the number of clusters; 
Figure 
\ref{F6a} corresponds to the experimental design just described; 
Figure \ref{F6b} corresponds to the experimental design with the same 
coefficients as the first one but $K=5$, $n=3$ and $500$ draws; 
Figure \ref{F6c} corresponds to a design with $K=5$, $n=3$ and $500$ draws, 
with 
coefficients $\boldsymbol{e}_1^T=(1/2~ 1~ 0~ 0~ 0), \boldsymbol{e}_2^T=(0~ 1~ 
1/2~ 0~ 0), \boldsymbol{e}_3^T=(0~ 0~ 1/2~ 1~ 0), \boldsymbol{e}_4^T=(0~ 0~ 0~ 
1~ 1/2), \boldsymbol{e}_5^T=(0~ 1~ 0~ 1~ 0).$
Finally, in Figure \ref{F6d} we set $K=10$, $n=15$ and $500$ draws but the 
coefficients $\boldsymbol{e}_i$ are drawn from a Dirichlet distribution with 
parameters $(.2,.2,.2,.2,.2)$. All these curves are decreasing and the speed of 
decrease slows down after the true number of clusters, even when the 
signals involved in each experiment are different. This ``elbow'' that seems to 
appear with the true number of clusters can be used as a empirical criteria to 
decide the number of clusters. Analogous results were obtained with several 
different simulation schemes.
Similar criteria are frequently used in cross 
validation methods. In this sense we propose this empirical criteria to get the 
number of possible clusters in real data analysis.

\subsubsection{Bootstrapping} 
The second option is based on a bootstrap 
procedure to approximate the distribution of the TV distance between 
estimated spectral densities, that was proposed in \cite{Euan16} and 
will be reported in detail in a different manuscript. Here, we will use this 
methodology to approximate the distribution of the total 
variation distance between two clusters. Then, we use this 
approximated distribution to choose a threshold 
for the TV distance between estimated spectra to decide whether or not the 
clusters should be merged.

The algorithm proposed in \cite{Euan16} to obtain a bootstrap sample of the TV 
distance is the following. 
\begin{enumerate}
 \item From $X_1(t)$ and $X_2(t)$, estimate $\widehat{f}^{X_1}(\omega)$ and 
$\widehat{f}^{X_2}(\omega)$ and take $ \displaystyle 
\widehat{f}(\omega)=
\frac{\widehat{f}^{X_1}_N(\omega)+\widehat{f}^{X_2}(\omega)}{ 2 }.$
 \item \label{A3} Draw $Z(1),\ldots,Z(T)\sim N(0,1)$ i.i.d random variables, 
then estimate $\hat{f}^Z_N(\omega)$ using also the smoothed periodogram.
 \item \label{A4} The bootstrap spectral density will be $\hat{f}_N^B(\omega)= 
\hat{f}_N(\omega)\hat{f}^Z_N(\omega).$ 
 \item Repeat \ref{A3} and \ref{A4} and estimate $\hat{d}_{TV}$ using the 
bootstrap spectral densities, i.e., 
$$d_{TV}^B= d_{TV} (\hat{f}_N^{B_1},\hat{f}_N^{B_2}),$$
where $\hat{f}_N^{B_i}$, $i=1,2,$ are two bootstrap spectral densities 
using different replicates of the process $Z(\cdot)$. 
\end{enumerate}
The bootstrap spectral density presented in \ref{A3} and \ref{A4} is motivated 
by the method presented in \cite{Paparoditis15}. \cite{Euan16} shows that 
this procedure produces a consistent estimation 
of the distribution of the TV distance. To extend this procedure to choose the 
number of clusters, first we need to note that due to the hierarchical 
structure of the algorithms used in all the methods proposed, the test 
following test are equivalent:
\begin{align*}
 &H_0: ~k-1 ~\mbox{Clusters} \qquad\mbox{vs} \qquad H_A: ~k ~\mbox{Clusters}, \\
 &H_0: ~1 ~\mbox{Cluster} \qquad\mbox{vs} \qquad H_A: 2 ~\mbox{Clusters},
\end{align*}
since the $(k-1)$ clusters are built by joining two of the $k$ 
clusters. 

In addition, we will use this option in the method presented 
by \cite{Alvarez15} to select the number of clusters and compare with the HSM 
method. The distribution of the total variation distance between two clusters 
depends on the clustering procedure. When using the HSM method we aim to 
approximate the 
distribution of the distance between the mean spectra in each cluster while for 
the hierarchical clustering with the TV distance, we need to produce samples 
from each cluster to approximate the distribution of the distance calculated 
through the link function.

The procedure of this test will be:
\begin{itemize}
 \item Run the clustering procedure, either the HSM method or hierarchical 
clustering with average or complete linkage. 
 \item Identify the two clusters that are joined to get the $(k-1)$ clusters.
 \item Under the null hypothesis where the two clusters should be merged, 
denote the common representative spectral estimate to be $\widehat{f}$.
 \item Simulate the spectra with the bootstrap procedure to compute the 
TV 
distance. We consider two cases:
 \begin{itemize}
  \item[\textit{Case 1.}] When using the HSM method simulate two spectral 
densities from the common spectra $f$ and compute the TV distance between them. 
Repeat this procedure $M$ times.
  \item[\textit{Case 2.}] When using hierarchical clustering with the TV 
distance simulate two sets of spectral densities of size $g_1$ and $g_2$ from 
the common spectra $f$, where $g_i$ are the number of members in cluster 
$i=1,2$ (clusters to be joined). Compute the link function (complete or 
average) between these two sets of spectra using the TV distance. 
distance.    
 \end{itemize}
 \item Run the test with the bootstrap sample.
\end{itemize}
\textit{\bf{Remark.}} Notice that this test assumes that there exits a common 
spectra 
$f$ within each cluster.  In practice, it is possible that the spectra in one 
cluster could vary slightly and in that setting we could cast this under mixed 
effects model (\cite{Krafty11}).

To investigate the performance of this procedure, we used \textbf{Experiments 
1} 
and \textbf{2}. We considered the TV distance to feed a hierarchical algorithm 
with two different link functions, namely, average and complete. Also, we 
considered the 
HSM method with the average version. In this case, we use $N=500$ replicates 
for each experiment.

Tables \ref{Exp1NC} and \ref{Exp2NC} present the proportion of times that the 
null hypothesis is rejected. To reject we used the bootstrap quantile of 
probability $\alpha$. We did not expect to have a proportion of rejection equal 
to $\alpha$, since in the case of using the complete or average link, these 
values are not a direct observation of the TV distance.
However, we expect to have a good performance. In general, it could be possible 
to overestimate the number of clusters. 

\begin{table}\footnotesize
\centering
\begin{tabular}{ccccc}
Experiment 1 & \\
\hline \hline\\
Test &$\alpha$ & Complete& Average& HSM\\
\hline \\
1 cluster vs 2 clusters& 0.01&	1&	1&	1\\
      & 0.05&	1&	1&	1\\
      & 0.1&	1&	1&	1\\
2 clusters vs 3 clusters&	0.01&	0.052&	0.154&	0.008\\
&	0.05&	0.206&	0.492&	0.058\\
&	0.1&	0.382&	0.670&	0.164\\
\end{tabular}
\caption{Proportion of times that the null hypothesis is rejected. Complete 
corresponds to the TV distance in hierarchical algorithm with the complete link 
function, Average with the average link, and HSM is the 
hierarchical spectral merger with the average version.}\label{Exp1NC}
\end{table}

In \textbf{Experiment 1} the true number of clusters is $2$. From Table 
\ref{Exp1NC}, we observe that all methods rejected the hypothesis of one 
cluster, 
at all the significance levels. This means that the procedures did not 
underestimate the number of clusters. To test $2$ versus $3$ clusters, the 
proportion of 
rejection 
is high when we used the average link function, except in the case 
$\alpha=0.01$.
If we used the complete link, the results are better.
However, the best results are given by the HSM method.

\begin{table}\footnotesize
\centering
\begin{tabular}{ccccc}
Experiment 2 & \\
\hline \hline\\
Test &$\alpha$ & Complete& Average& HSM\\
\hline \\
2 clusters vs 3 clusters&	0.01&	0.968&	1&	0.25\\
      & 0.05&	1&	1&	0.924\\
      & 0.1&	1&	1&	0.998\\
3 clusters vs 4 clusters&	0.01&	0.072&	0.18&	0.002\\
&	0.05&	0.228&	0.924&	0.050\\
&	0.1&	0.376&	0.998&	0.106\\
\end{tabular}
\caption{Proportion of times that the null hypothesis is rejected, in 
Experiment 
2.}\label{Exp2NC}
\end{table}

\begin{figure}
 \centering
\subfigure[\label{PVTE1}]{\includegraphics[scale=.35]
{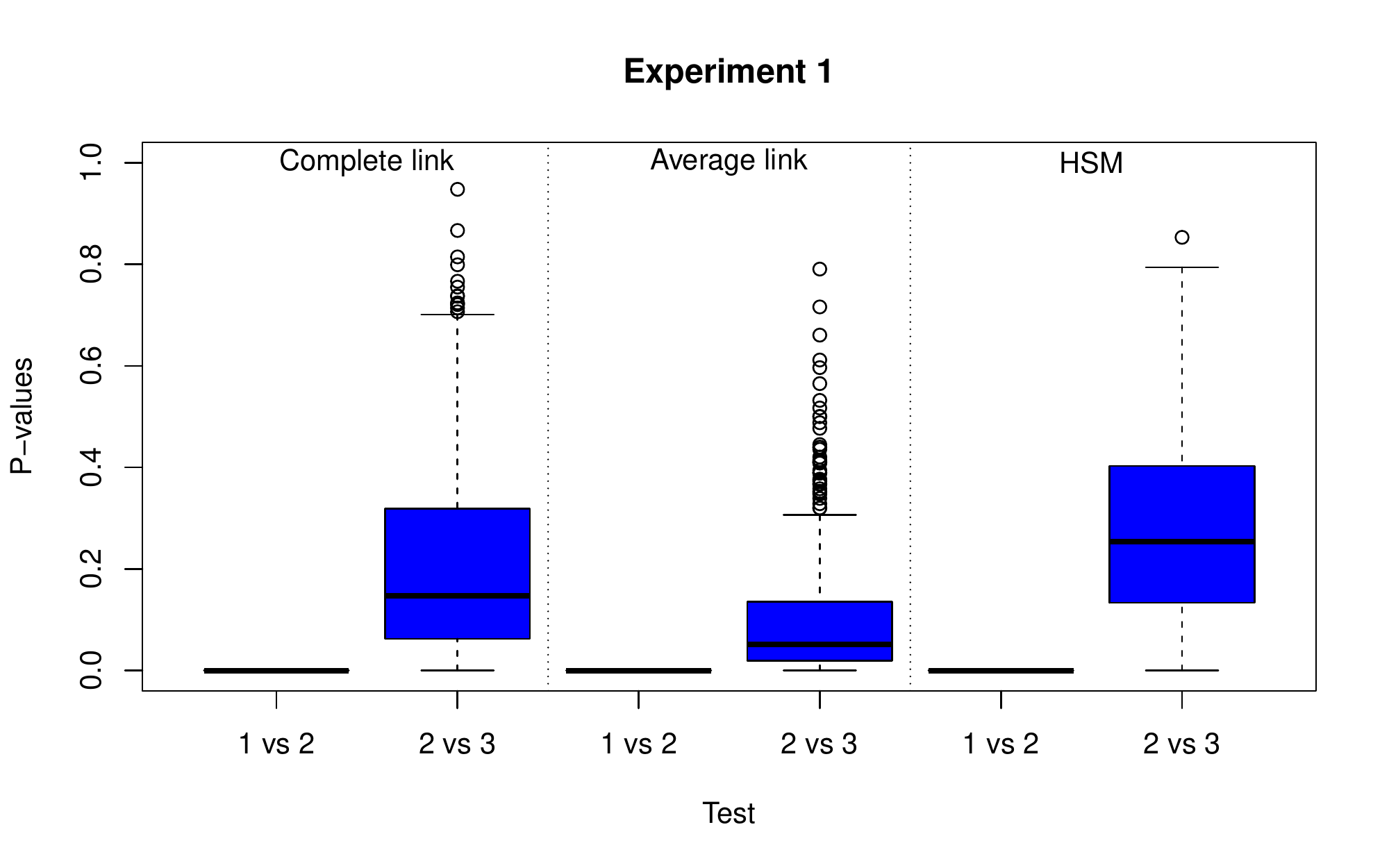}}
\subfigure[\label{PVTE2}]{\includegraphics[scale=.35]
{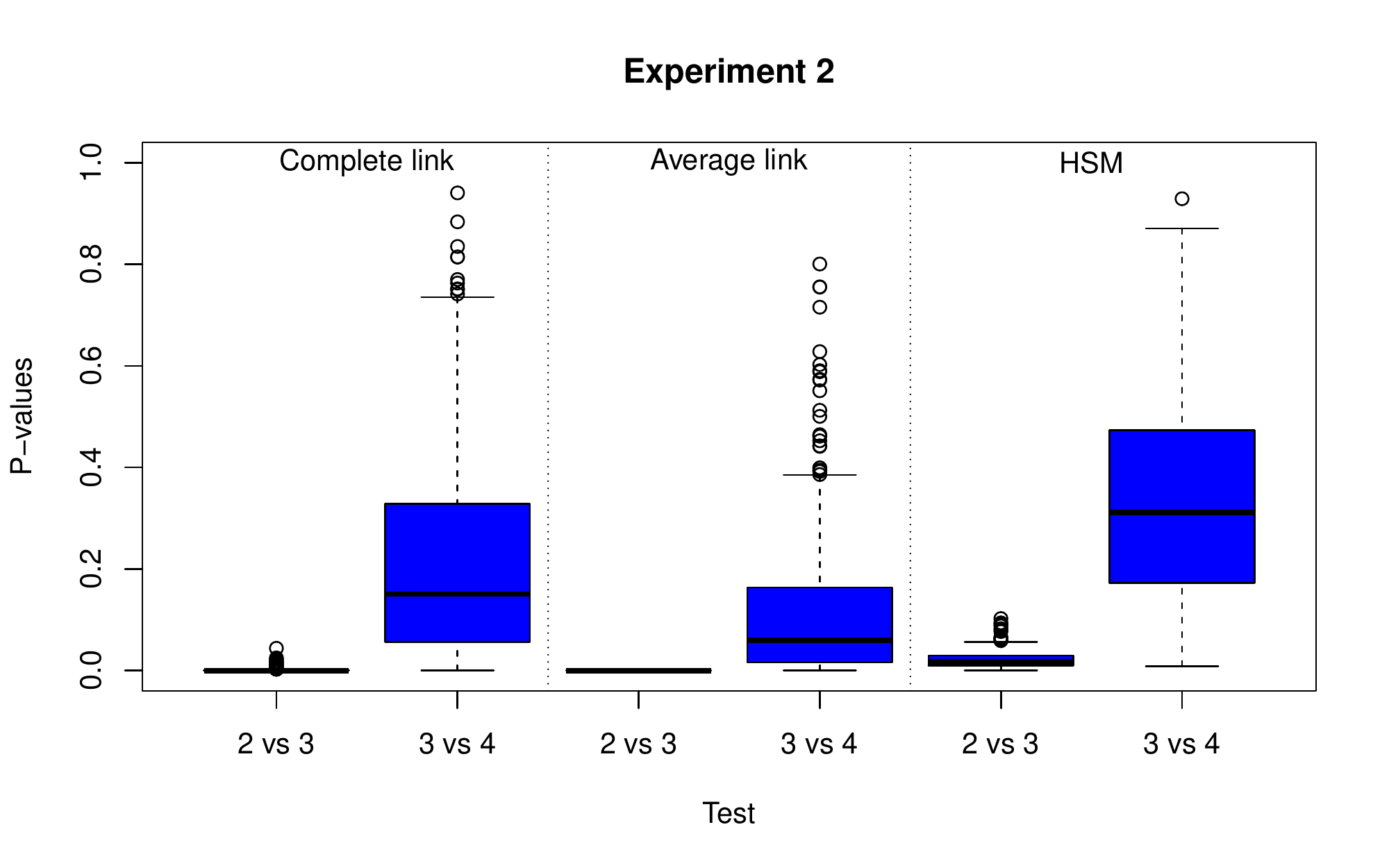}}
\caption{P-values obtained in the test of number of clusters using bootstrap 
samples.}\label{PvaluesTest}
\end{figure}

In \textbf{Experiment 2} the true number of clusters is $3$. This is a more a
difficult case, since the spectra are very close. From Table \ref{Exp2NC} when 
testing 2 versus 3 clusters, we observe that the complete and average link 
functions 
did not underestimate the number of clusters. However, the HSM method did not 
distinguish 3 clusters at a level $\alpha=0.01$, but the results are better at 
higher 
levels. For testing 3 versus 4 clusters, HSM performed the best, 
followed by the complete 
link. Again it was necessary to have a small value of $\alpha$ in order that 
the average 
link gave a reasonable performance.

Figure \ref{PvaluesTest} shows the p-values obtained comparing the value 
from each simulation with the bootstrap distribution. We confirm the 
fact 
that the underestimation of the number of clusters has low probability, almost 
zero in some cases, for the three methods. When the number of clusters to test 
is the correct one, $2$ in \textbf{Experiment 1} and $3$ in \textbf{Experiment 
2}, the p-values are widely distributed in the case of the complete link and 
HSM method. With the average link, the p-values are smaller compared to the 
other methods. 
In general, this test has a good performance when one uses the complete link or 
the HSM method.
\section{Analysis of the ocean waves and EEG data}
We developed the \textit{HSMClust} package written in R \nocite{RR} that 
implements our proposed clustering method. The package can be downloaded from
\url{http://ucispacetime.wix.com/spacetime#!project-a/cxl2}.

{\bf Example 1: EEG Data.} Our first data example is the resting-state EEG data 
from a single subject. The goal here is to cluster 
resting-state EEG signals from different channels that are spectrally 
synchronized, i.e.,
that show similar spectral profiles. This subject is from the healthy 
population 
and the EEG clustering here will serve as a 
``standard" to
which the clustering of stroke patients (with severe motor impairment) will be 
compared. Data were collected at 1000 Hz and pre-processed in the following 
way. The continuous EEG 
signal was low-pass filtered at 100 Hz, segmented into non-overlapping 1 second 
epochs, and detrended. The original number of channels 256 had to be reduced to 
194 because of the presence of artifacts that could not be corrected (e.g., 
loose leads). 

\begin{figure}
\centering
\includegraphics[scale=.3]{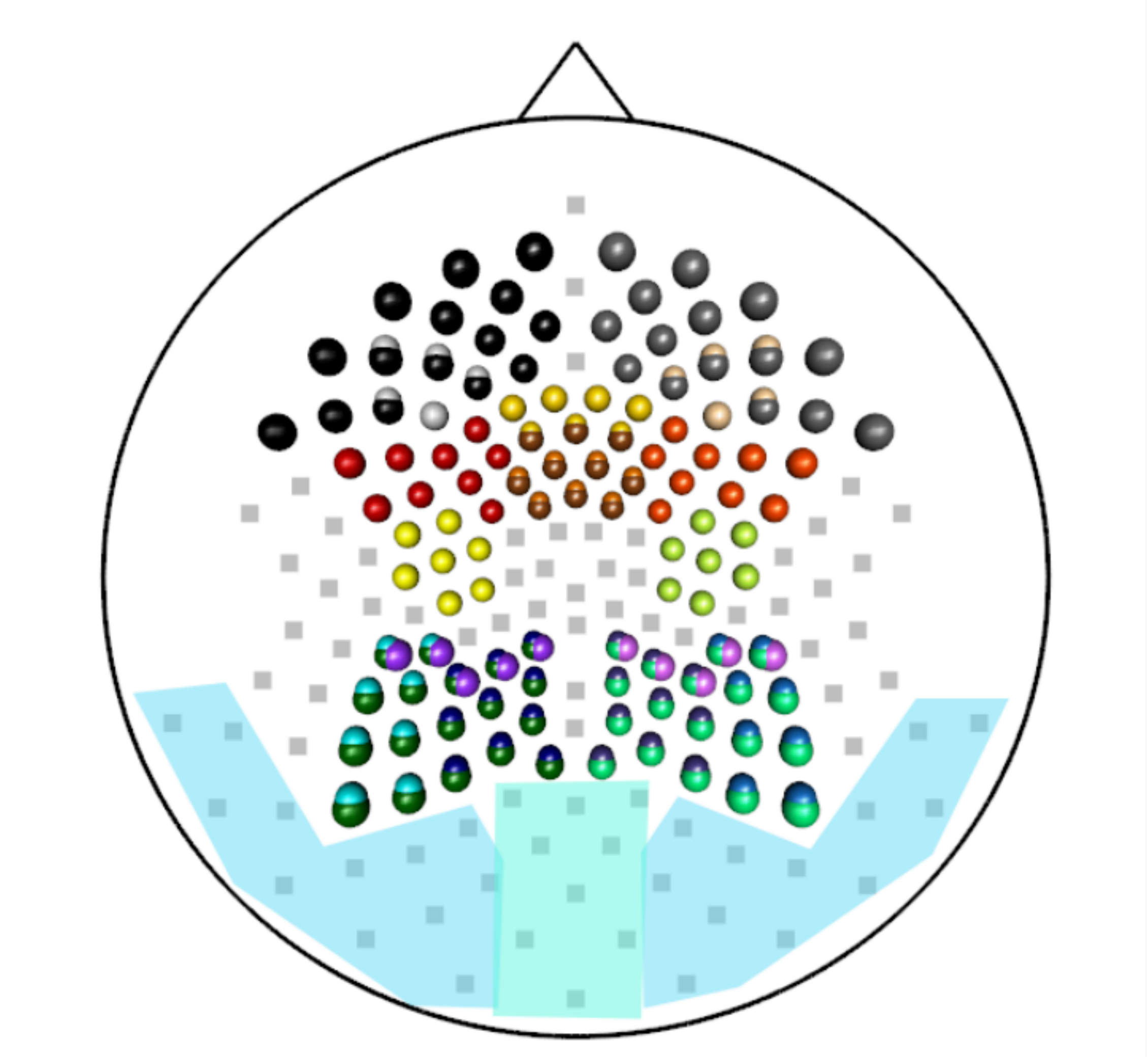}
\includegraphics[scale=.3]{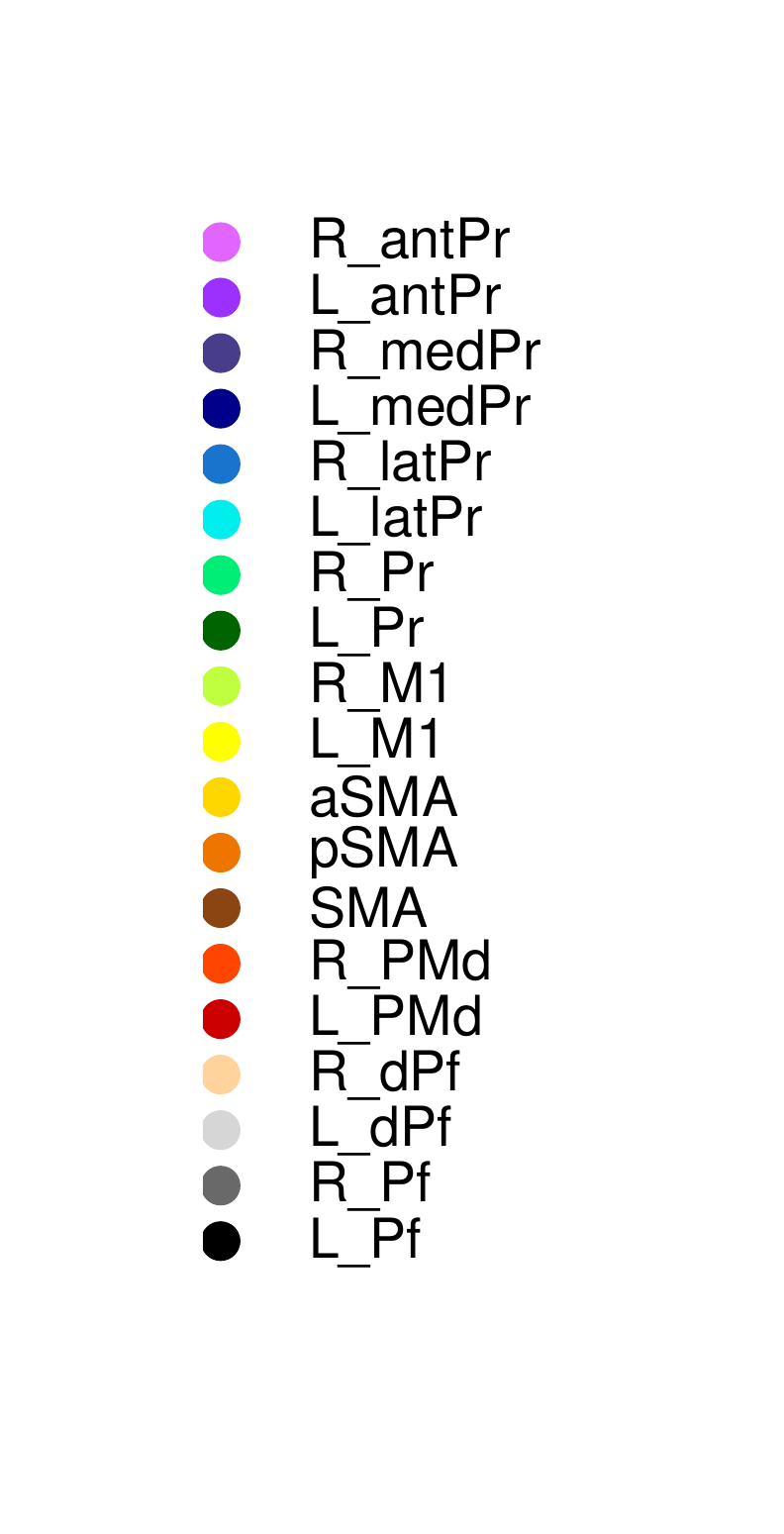}
\caption{Brain regions: Left/Right Prefrontal (L\_Pf, 
R\_Pr), Left/Right Dorsolateral Prefrontal (L\_dPr, R\_dPr), Left/Right 
Pre-motor (L\_PMd, R\_PMd), Supplementary Motor Area (SMA), anterior SMA 
(aSMA), 
posterior SMA (pSMA), Left/Right Primary Motor Region (L\_M1, R\_M1), 
Left/Right 
Parietal (L\_Pr, R\_Pr), Left/Right Lateral Parietal (L\_latPr, R\_latPr), 
Left/Right Media Parietal (L\_medPr, R\_medPr), Left/Right Anterior Parietal 
(L\_antPr, R\_antPr). Gray squared channels do not belong to any of these 
regions; Light blue region corresponds to right and left occipital and light 
green region corresponds to central occipital.}\label{F1}
\end{figure}

The EEG channels were grouped into $19$ pre-defined regions in the brain as 
specified in \cite{Wu14}: prefrontal (left-right), dorsolateral prefrontal 
(left-right), pre-motor (left-right), supplementary motor area (SMA), anterior 
SMA, posterior SMA, primary motor region (left-right), parietal (left-right), 
lateral parietal (left-right), media parietal (left-right) and anterior 
parietal 
(left-right). Figure \ref{F1} shows the locations of these regions on the 
cortical surface. 

We present the results for subject BLAK at epoch 25 of the 
resting state. The interpretations will employ 
the usual division in frequency ranges for the analysis of spectral densities 
of EEG data: Delta (0.1-3 Hertz), Theta (4-7 Hertz), Alpha (8-12 Hertz), Beta 
(13-30 Hertz) and Gamma (31-50 Hertz). 
\begin{figure}
 \subfigure[\label{E21}]{\includegraphics[scale=.37]{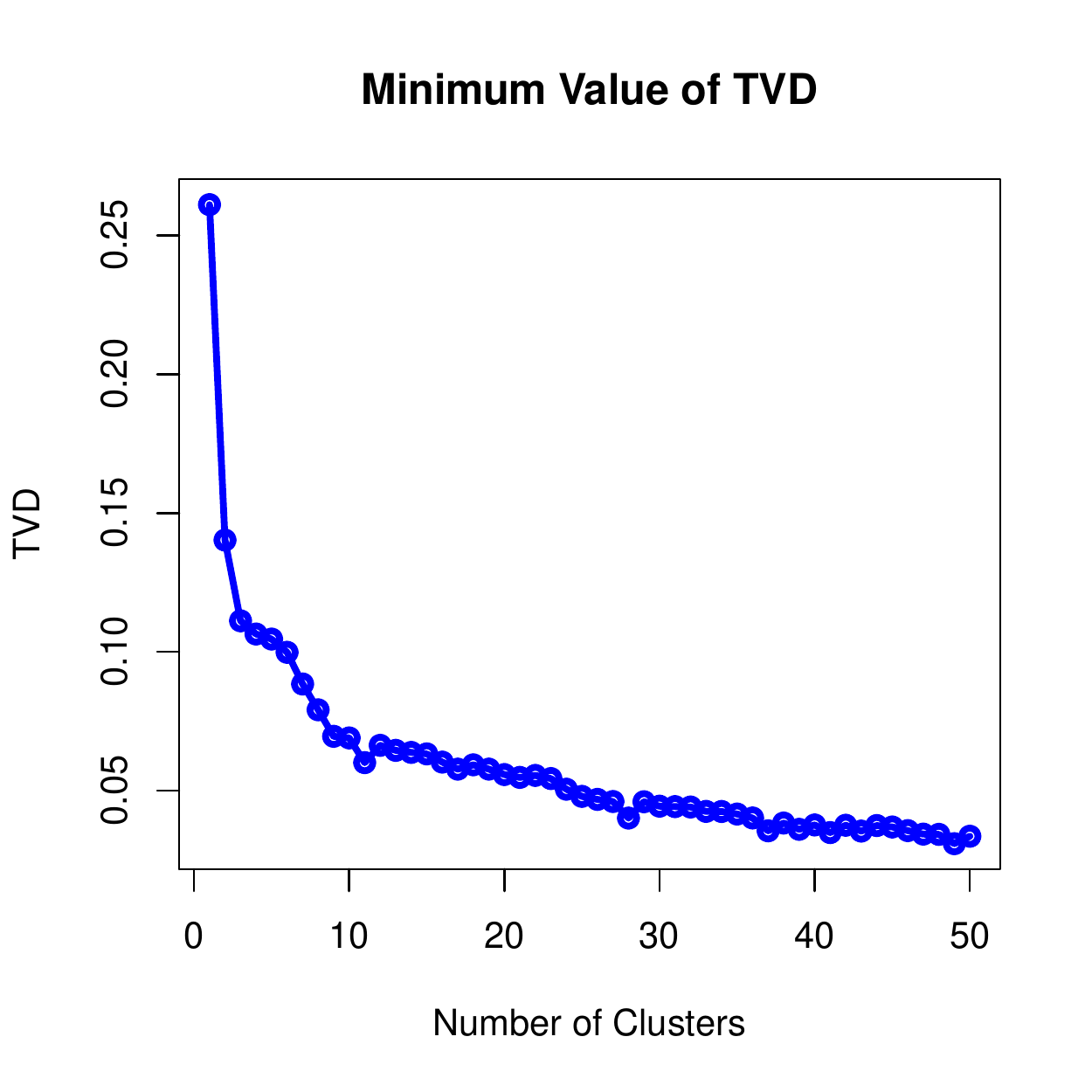}}
 \subfigure[\label{E22}]{\includegraphics[scale=.4]{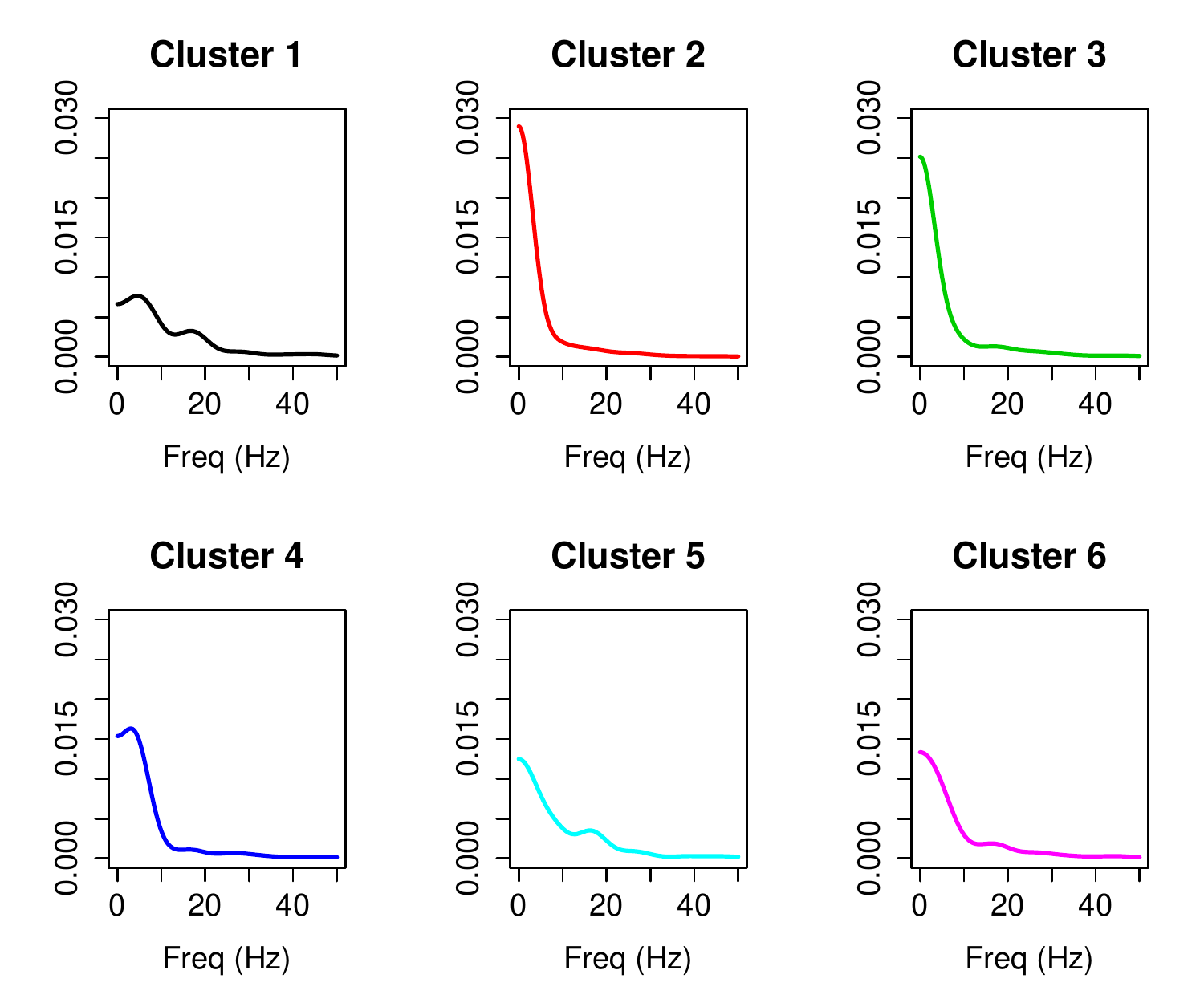}}
 \subfigure[\label{E23}]{\includegraphics[scale=.16]{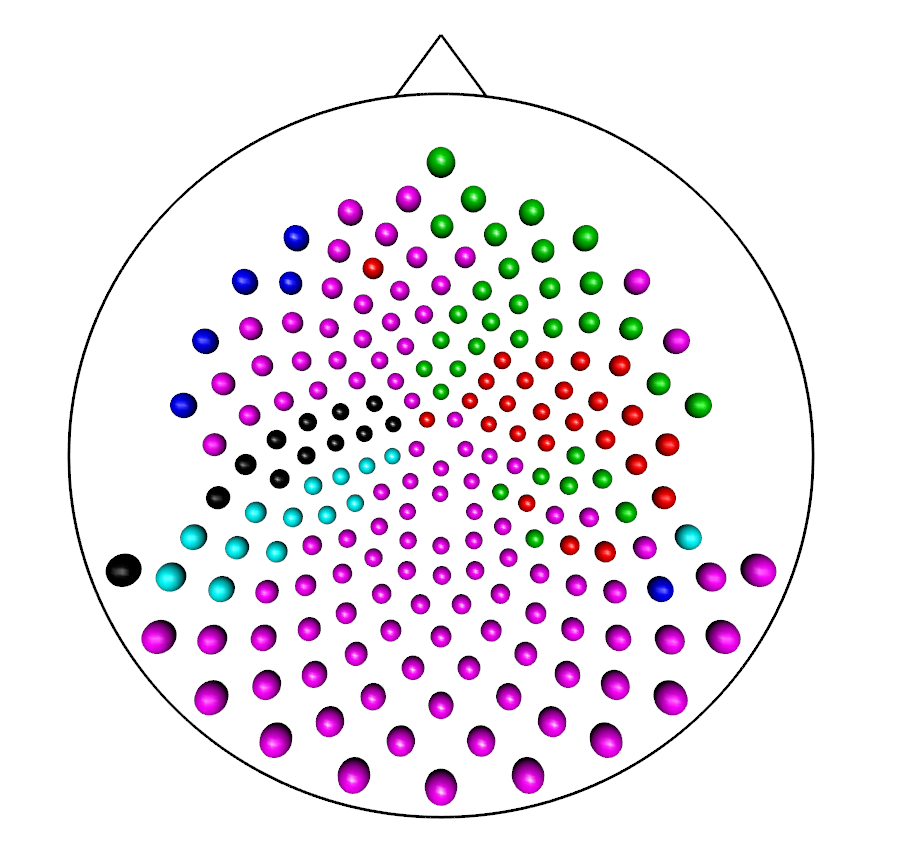}}
 \caption{Results using the \textit{HSM} function. (a) The trajectory if the 
minimum value of the TV distance, (b) Spectrum using all signals at each 
cluster and (c) Location of the clusters on the scalp.}
\end{figure}

Figure \ref{E21} shows the minimum value of the TV distance, in this case the 
``elbow'' appears around $6$ clusters. We analize the results for 7 
clusters as well, in this case the channels in two of the clusters were 
grouped into one since there were no significant evidence to reject 6 clusters 
over 7. Figure \ref{E22} and \ref{E23} show the shape of the mean normalized 
spectra by 
cluster and the location at the cortical surface. Many of the channels at the
occipital and left premotor regions belong to cluster $6$ (pink), 
which is dominated by frequencies at the theta and alpha bands (4-12 Hz). 
Cluster $2$ (red) is conformed by the channels at the right premotor region and 
this 
cluster is only influenced by the delta band (0-4 Hz). Clusters 1 (black) and 
 5 (sky blue) are the only ones with influence of the beta band and they 
are located at the left motor and left anterior parietal regions. 
 
The HSM method captures the behavior of the EEG during the resting state 
through 
the cluster membership of the EEG channels. In addition, the HSM method also 
identifies the frequency bands
that primarily drive the formation of the clusters. The clusters produced are
consistent for the most part with the anatomical-based parcellation of the 
cortical surface
and thus cluster formation based on the spectra of the EEGs could be used to 
recover the
spatial structure of the underlying brain process

\bigskip

{\bf Example 2. Sea Waves Data.} As a second example we consider wave-height 
data from the Coastal Data 
Information Program (CDIP) buoy 160 (44098 for the National Data Buoy Center) 
situated off the coast of New Hampshire, USA, at a water depth of 76.5 m., when 
a storm took place in December 2010.  
The data correspond to a period of 69 hours, divided into 138 intervals of 30 
minutes and recorded with a frequency of 1.28 Hz, between the 26th and the 29th 
of December.

The use of stochastic processes for the analysis and modeling of sea waves has 
a 
long history, starting with the work of \cite{Pierson}  and \cite{l-h1}. A 
model commonly used is that stationary Gaussian random process. For this 
particular sea waves data, the presence of a storm implies that the sea 
surface will be 
very unstable and the hypothesis of stationarity will only be valid for 
short periods of time. 

\begin{figure}
\centering
\includegraphics[scale=.3]{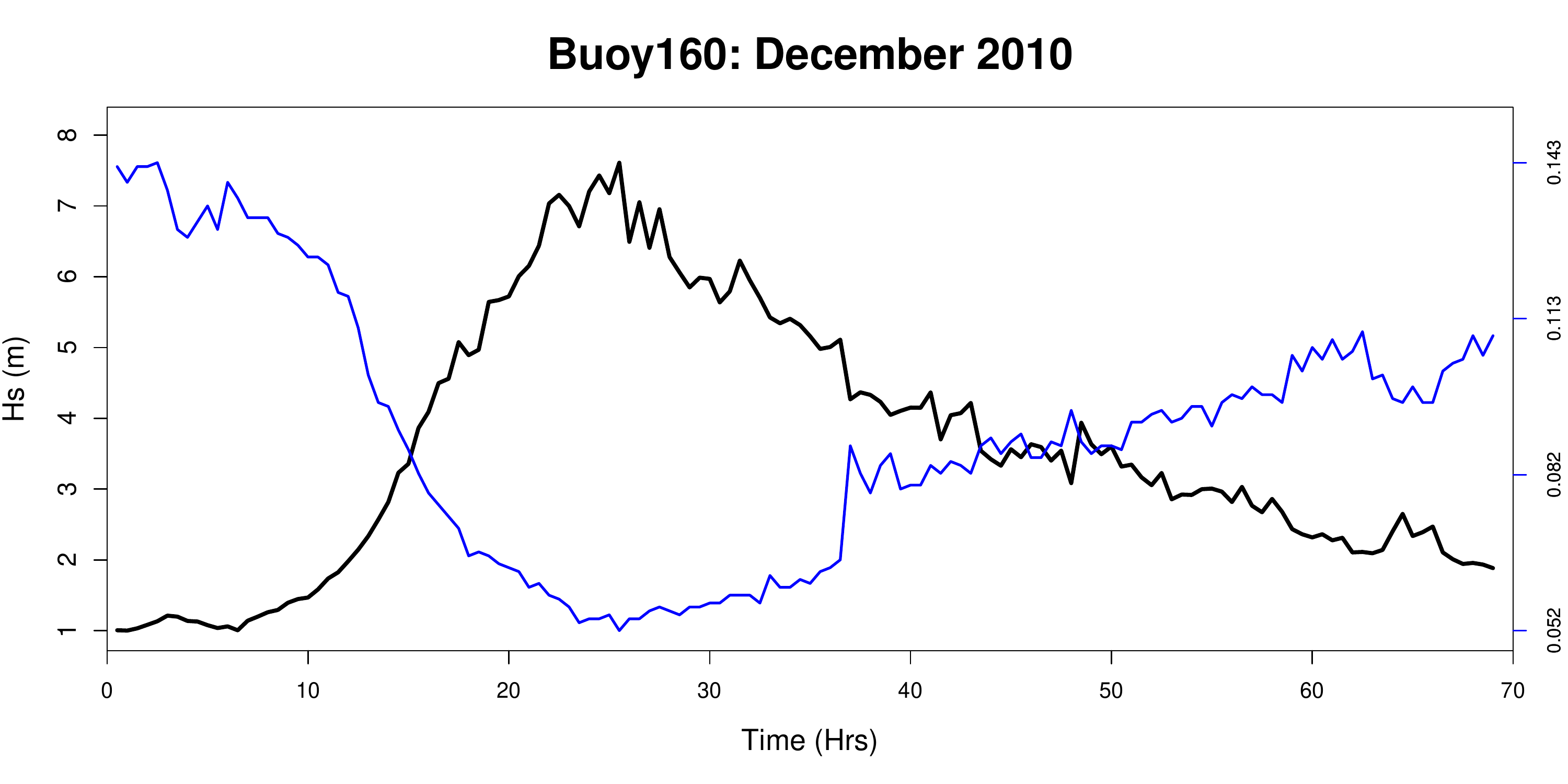}
\includegraphics[scale=.3]{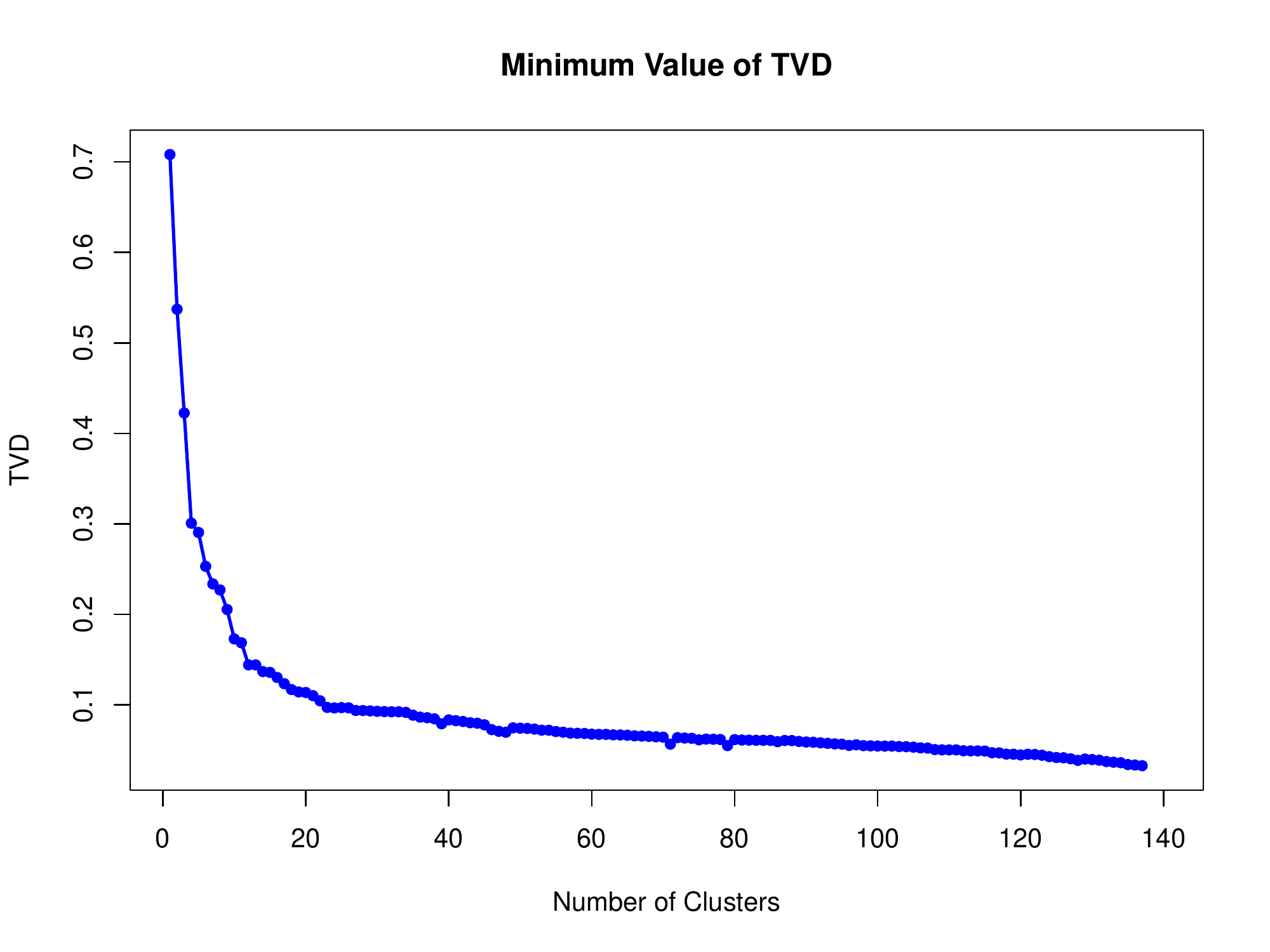}
\caption{Significant wave height (black) and peak frequency (blue) for the 69 
hour period (left). Trajectory for the minimum TV distance 
(right).}\label{F-olas1}
\end{figure}

Sea-height data from buoys are usually divided into short intervals, between 20 
and 30 minutes of duration, which are considered to be long enough to get a 
reasonable estimation of the spectral density, and short enough for stationarity 
to hold. For each interval, several numerical parameters are calculated from the 
estimated spectral density, which give a summary of the sea conditions. Two very 
important ones are the significant wave height and the peak frequency, or 
equivalently, the peak period. The former is a standard measure of sea severity 
and is defined as four times the standard deviation of the time series. The 
latter is just the modal spectral frequency, or the corresponding period. Figure 
\ref{F-olas1} (left) presents the evolution of both parameters for the time 
interval being considered.  

In \cite{Alvarez15} a segmentation method for long time series based on a 
clustering procedure was proposed and applied to a similar problem, but in a 
different location and different sea conditions. We will apply the same method 
with the clustering 
procedure introduced in this work. The main idea is that the spectral density 
should be able to identify the time intervals coming from the same stationary 
process. Thus, since the data are already divided into 30-minute intervals, a 
clustering procedure on the corresponding spectral densities should identify 
intervals having similar oscillatory behavior. If these intervals are 
contiguous in time, then they are considered part of a longer stationary 
interval. More details on the procedure can be found in  \cite{Alvarez15}. 
 
\begin{figure}
\centering
\includegraphics[scale=.3]{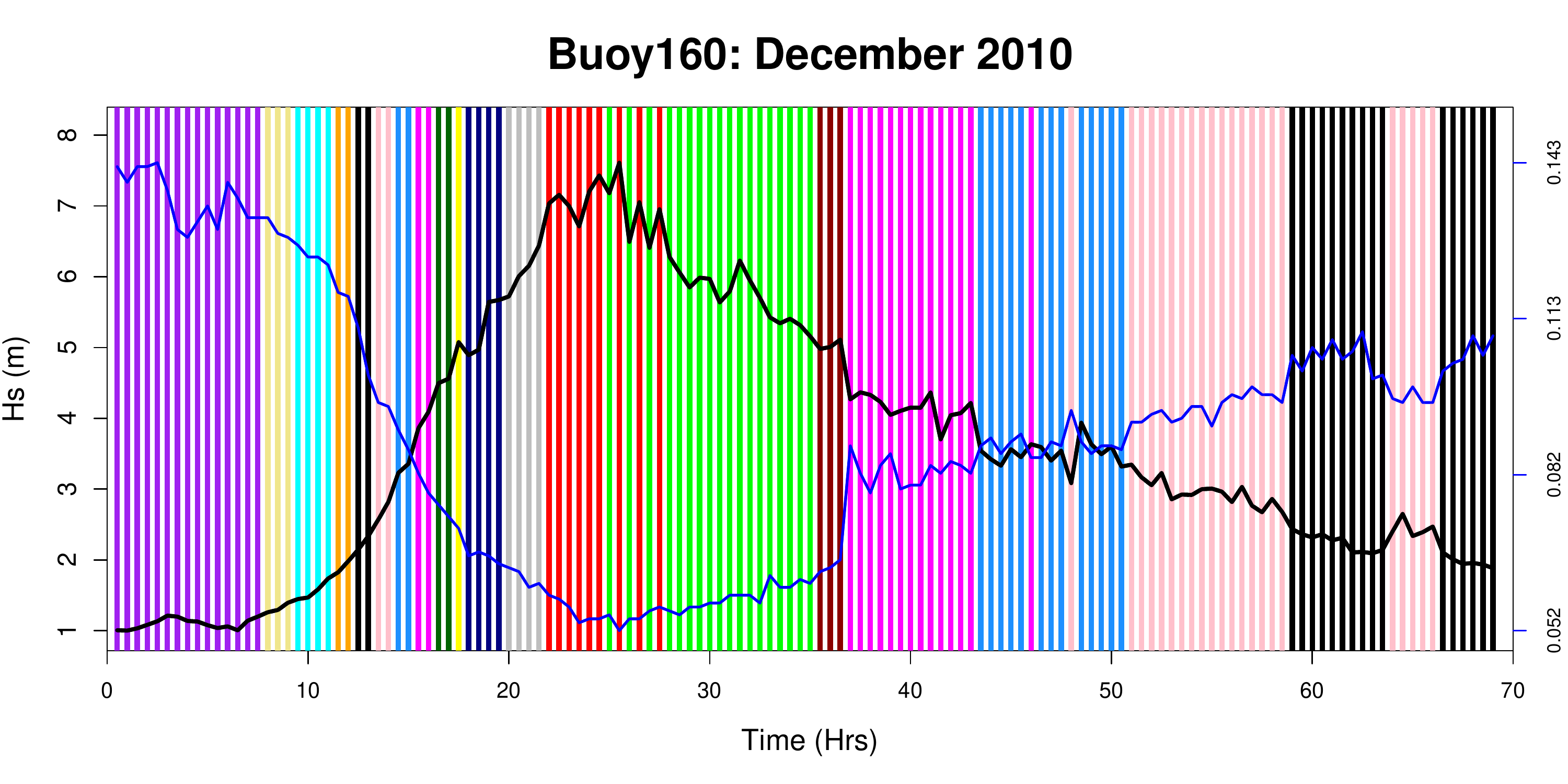}
\caption{Segmentation for the 96 hour period. Different colors correspond to 
different clusters.}\label{F-olas2}
\end{figure}

As in the previous example, the trajectory for the minimum TV distance (Figure 
\ref{F-olas1}, right) is used to get a first estimate for the number of 
clusters, which in this case is 15. A bootstrap test supports this choice 
(p-value = 0.01). The segmentation resulting from this clustering is shown in 
Figure \ref{F-olas2}, where the different clusters correspond to different 
colors in the vertical lines.

It is interesting to compare these results with a similar analysis of a 
different data set in \cite{Alvarez15}, event thought the clustering procedures 
differ. In that case, for a time series of 96 hours recorded in Waimea Bay, 
Hawaii in January, 2003, only 5 or 6 clusters were detected, depending on the 
linkage function (average or complete) used for the hierarchical agglomerative 
procedure. During the period studied for the Hawaii data there is also a storm, 
but of a much smaller magnitud, with a significant wave height always below 5 m. 
In contrast, for the data considered in this paper, significant wave height 
surpasses 7 m. at the peak of the storm. The intensity of the storm is a 
possible factor in the presence of more clusters of shorter duration.
Table \ref{T-olas1} gives the duration of consecutive intervals within a common 
cluster, with a mean value of 2.23 h and a median of 1.5 h. 
\begin{table}\footnotesize
\centering
\begin{tabular}{ccccccccccc}
Time (h)                   & 0.5 & 1 & 1.5 & 2 &	2.5 & 3 &	5 & 6.5 
& 7.5 & 8\\
Number of intervals &  9   & 6 &  3  &  3 &	3 & 2 & 1 & 1 & 2 & 1\\
\end{tabular}
\caption{Frequency table for the duration of consecutive intervals within a 
common cluster.}\label{T-olas1}
\end{table}

\begin{figure}
\centering
\includegraphics[scale=.5]
{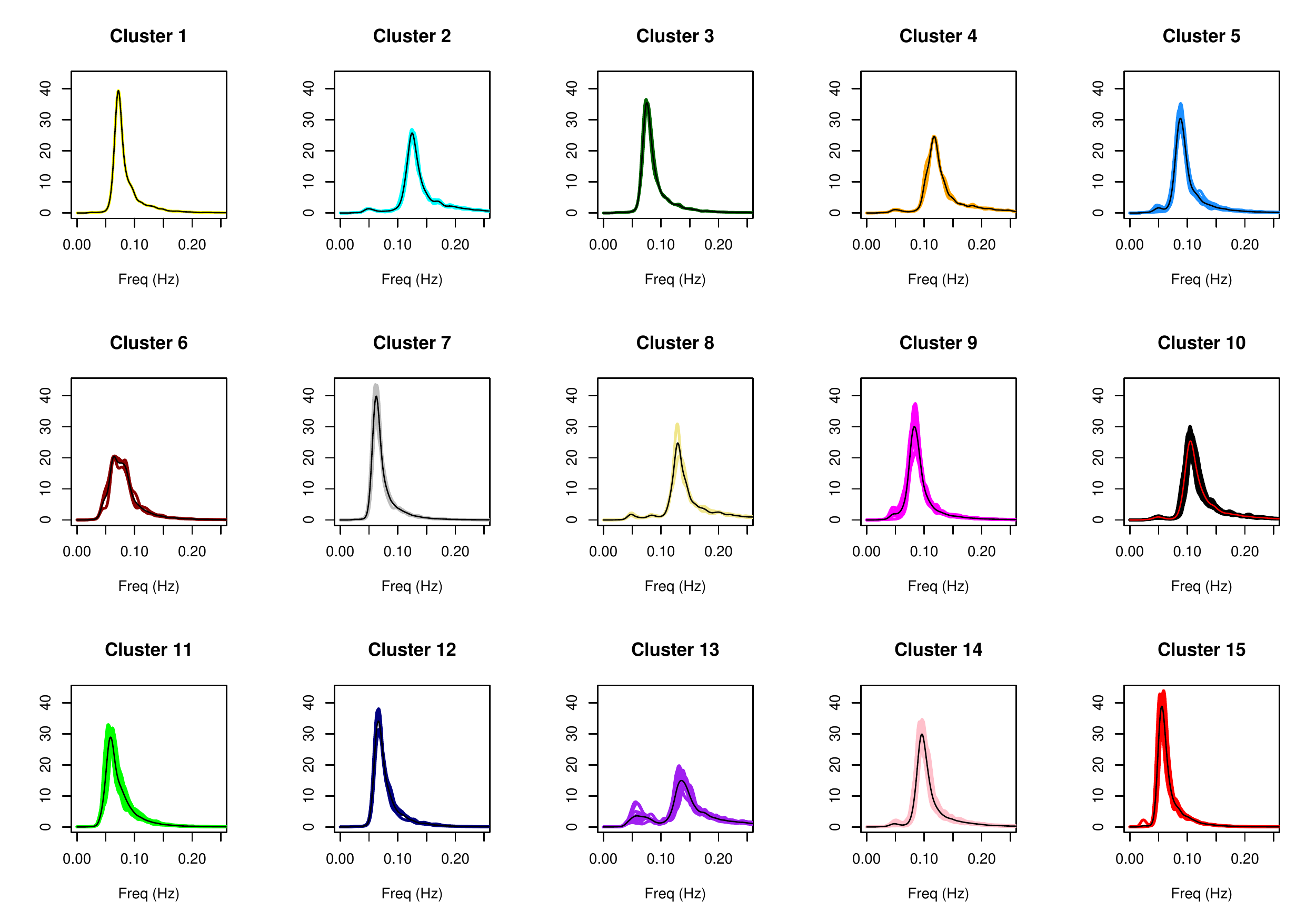}
\caption{Normalized spectral densities for the 15 clusters, the mean spectral 
density is represented in black.}\label{F-olas3}
\end{figure}

Figure \ref{F-olas3} presents the normalized spectral densities for the 15  
clusters. The graphics show that  bimodal spectral densities characterize some 
clusters, even though in some cases the secondary mode is much smaller than the 
main one. Bimodal spectra correspond to the presence of a second train of waves 
with different dominating frequency, and the presence of this secondary train 
of 
waves, even if it is small, is captured by the method as an important feature 
of 
the spectra. In other cases there are differences on the dispersion of the 
spectral density or in the location of the modal frequency. The procedure 
employed yields a spectral characterization of the different clusters, which 
can 
be linked to the statistical characteristics of their duration, something which 
is useful for the design of marine structures and energy converters.

\section{Discussion and Future Work}
A new clustering procedure for time series based on the spectral densities and 
the total variation (TV) distance, the Hierarchical Spectral Merger (HSM) 
algorithm, was introduced. Using numerical experiments, the proposed HSM method 
 was compared with other available procedures, and the results show that its 
performance comparable to the best, in addition to having an intuitive 
interpretation. Applications to two data from different scientific fields were 
also presented, showing that the method has wide applicability in many different 
areas.

However, the method is not free from limitations and further developments are needed.
The fact that each cluster has a characteristic spectral density, with respect to which all distances are measured, may provide a way of correcting classification mistakes that occur due to the evolution of the clusters. This methodology would give a more robust clustering method.

\section{Acknowledgements}
The software WAFO developed by the Wafo group at Lund University of
Technology, Sweden, available at
http://www.maths.lth.se/matstat/wafo was used for the calculation of
all spectral densities and associated spectral characteristics. The
data for station 160 were furnished by the Coastal Data Information
Program (CDIP), Integrative Oceanographic Division, operated by the
Scripps Institution of Oceanography, under the sponsorship of the
U.S. Army Corps of Engineers and the California Department of
Boating and Waterways (http://cdip.ucsd.edu/).

This work was partially supported by 1) CONACYT, M\'exico, scholarship AS 
visiting research student, 2) CONACYT, M\'exico, Proyectos 169175 An\'alisis 
Estad\'istico de Olas Marinas, Fase II and 234057 An\'alisis Espectral, Datos 
Funcionales y Aplicaciones, and 3) Centro de Investigaci\'on en Matem\'aticas 
(CIMAT), A.C. Eu\'an and Ortega wish to thank Prof Pedro C. Alvarez Esteban for 
several fruitful conversations on the methodology proposed on this paper. C. 
Eu\'an wishes to thank to UC Irvine Space Time Modeling Group for the invitation 
to collaborate as a visiting scholar in their research group.

The research conducted at the UC Irvine Space-Time Modeling
group (PI: Ombao) is supported in part by the National Science
Foundation Division of Mathematical Sciences and the Division
of Social and Economic Sciences. The authors thank Dr. Steven
C. Cramer of the UC Irvine Department of Neurology for sharing
the EEG data that was used in this paper.
This work was done while J.O. was visiting, on sabbatical leave from CIMAT and with
support from CONACYT, M\'exico, the Departamento de Estad\'istica e I.O., Universidad de
Valladolid. Their hospitality and support is gratefully acknowledged.


\end{document}